\documentclass[reprint]{revtex4-2}
\usepackage{graphicx}
\usepackage{amsmath}
\usepackage{amssymb}
\usepackage{slashed}
\usepackage{hyperref}
\usepackage{enumerate}
\usepackage{xcolor}
\usepackage{tabularx}

\usepackage{csvsimple}

\graphicspath{ {./Images/} }

\begin{document}

\title{Hybrid Star Properties with NJL and MFTQCD Model: A Bayesian Approach}

\author{Milena Albino}
\email{milena.albino@student.uc.pt}
\affiliation{Department of Physics, CFisUC, University of Coimbra, P-3004 - 516 Coimbra, Portugal}

\author{Tuhin Malik}
\email{tm@uc.pt}
\affiliation{Department of Physics, CFisUC, University of Coimbra, P-3004 - 516 Coimbra, Portugal}

\author{Márcio Ferreira}
\email{marcio.ferreira@uc.pt}
\affiliation{Department of Physics, CFisUC, University of Coimbra, P-3004 - 516 Coimbra, Portugal}

\author{Constança Providência}
\email{cp@uc.pt}
\affiliation{Department of Physics, CFisUC, University of Coimbra, P-3004 - 516 Coimbra, Portugal}

\date{\today}

\begin{abstract}
    \noindent
    \textbf{Context:}
    The composition of the core of neutron stars is still under debate.
    One possibility is that because of the high densities reached in their cores,
    matter could be deconfined into quark matter.
    
    \noindent
    \textbf{Aims:}
    In this work, we investigate the possible existence of hybrid stars, using  microscopic models to describe the different phases of matter. Within the adopted microscopic models we aim at calculating  both the properties of neutron stars and the properties of matter. In particular, we want to probe the influence of pQCD calculations and analyze the properties that identify a transition to deconfined matter.
    
    \noindent
    \textbf{Methods:}
    A Bayesian approach using  a Markov-Chain Monte Carlo
    sampling process is  applied to generate 8 sets of equations of state.
    A Maxwell construction is adopted to describe the deconfinement transition.
    For the hadronic phase, we consider a stiff and a soft EOS 
    obtained from the Relativistic Mean Field (RMF) model with non-linear meson terms.
    For the quark phase, we use two different models:
    the Nambu-Jona-Lasinio (NJL) model with multiquark  interactions
    and the Mean Field Theory of QCD, a model  similar to the MIT bag model with a vector term.
    Bayesian inference was applied to determine the model  parameters that satisfy
    the X-ray observations from NICER and have phase transition at densities between 0.15 and 0.40 fm$^{-3}$.
    We have also applied restrictions from the pQCD calculations to half of the  sets.

    \noindent
    \textbf{Results:}
    Hybrid stars are compatible with current observational data.
    The restrictions of pQCD  reduce the value of the maximum mass. However, even when applying this restriction,
    the models were able to reach values of $M_\text{max} = 2.1-2.3 M_\odot$. The conformal limit was still not attained at the center of the most massive stars. The vector interactions are essential to describe hybrid stars with a mass above 2$M_\odot$. The  multiquark interactions introduced may affect the limits of some quantities considered as indicators of the  presence of a deconfined phase. It is possible to find a set of EOS, that predict that inside NS the renormalized matter  trace anomaly is  always positive. 
\end{abstract}

\maketitle

\section{Introduction}
Neutron stars (NSs) represent some of the densest forms of matter observable in the universe, providing unique natural laboratories for studying the extreme states of quantum chromodynamics (QCD) \cite{Fukushima:2010bq,Bazavov:2011nk}. The cores of NSs, in particular, may host a myriad of exotic phases of matter, offering a window into the behavior of QCD under such extreme conditions \cite{glendenning,Rezzolla:2018jee}. Theoretical predictions based on perturbative QCD (pQCD) suggest the presence of deconfined quark matter at extraordinarily high baryon densities, approximately forty times greater than the density of nuclear matter at saturation $\rho_0$ \cite{pqcd}. The most challenging region to study theoretically is, however, at intermediate densities i.e. few times nuclear matter saturation
density which is actually relevant for the matter in the core of NSs \cite{glendenning}.

At these intermediate densities, fundamental calculations like Lattice QCD face significant computational obstacles, particularly the well-known sign problem in simulations at finite baryon densities. The current ability of chiral effective field theory ($\chi$EFT) to accurately compute the EOS at relatively low baryon densities, below twice saturation density \cite{Hebeler:2013nza,Drischler:2015eba} does not cover the range of densities that is hypothesized in neutron star (NS) cores. In contrast, various effective models that suggest a range of novel and exotic phases of quark matter in the intermediate density have been studied over the past decades. These include pion superfluidity \cite{Son:2000xc, Ebert:2005wr, Barducci:2004tt}, different types of color superconductivity such as two-flavor color superconductivity \cite{Alford:1997zt, Mishra:2003nr, Bonanno:2011ch, Abhishek:2021onm} and the color flavor locked (CFL) phase \cite{Alford:1998mk}, along with more intricate configurations like the Larkin-Ovchinkov-Fulde-Ferrel (LOFF) \cite{Mannarelli:2006fy, Rajagopal:2006ig} and crystalline superconductivity phases. To support the possibility of quark matter cores within neutron stars, recent model-independent studies combining astrophysical observations and theoretical calculations have shown that the interior properties of the most massive neutron stars align with the characteristics of a deconfined quark phase, suggesting the existence of sizable quark-matter cores in stars approaching or exceeding two solar masses \cite{Annala:2019puf,Altiparmak:2022bke,Somasundaram:2022ztm}.
However, no information on the composition was considered in the intermediate density regime.
The behavior of some quantities as the speed of sound, the polytropic index \cite{Annala:2019puf},
the renormalized trace anomaly $\Delta = 1/3 - P / \epsilon$  \cite{Fujimoto:2022ohj} and 
the measure of conformality $d_c \equiv \sqrt{\Delta^2 + \Delta'^2}$,
where $\Delta'$ is the logarithmic derivative with respect to the energy density \cite{Annala:2023cwx},
were analyzed and proposed as possibly indicating the existence of a quark core in NSs. These analyzes were carried out considering agnostic model independent descriptions of the EOS.

One of the aims of this work is to describe hybrid stars considering physically justified hadron and  quark models in order to study the possible existence of a quark core inside a NS.  The quantities that have been proposed to identify the presence of a quark core within agnostic models will be studied considering these microscopic hybrid EOS,  and confronted with the values proposed as indicative of the presence of deconfined matter.

Since the formulation of QCD, the possible existence of quarks inside neutron stars has been questioned \cite{glendenning}. One of the first models of hybrid stars \cite{BAYM1976241} predicted that
the quark-hadron transition would only happen around $10 \rho_0$, making the existence of hybrid stars very unlikely. However, as new models emerged, the results pointed to a  phase transition at lower densities, in which it was possible to have quarks inside neutron stars. Since then, various studies have been carried out, considering different hadronic models  \cite{Oertel:2016bki, constrain_apply, Malik_2022, Hebeler:2013nza}, hybrid models  \cite{Bonanno:2011ch,Logoteta:2013ipa,Benic:2014jia,renan_wr,Matsuoka:2018huq,Alvarez-Castillo:2016oln,PhysRevD.99.083014,Schramm_2016,parisi2020hybrid,PhysRevD.101.123029,PhysRevD.103.023001,PhysRevC.103.045205,Pfaff:2021kse,Xie:2020rwg,Ayriyan:2021prr,Takatsy:2023xzf}
or even model-independent studies \cite{Annala:2019puf,Altiparmak:2022bke,Fujimoto:2022ohj,Annala:2023cwx}.
Despite numerous studies and the rapid development of pulsar detection techniques, it is still not possible to determine with confidence the existence of quarks in the NS. In this work, we apply a Bayesian inference approach to generate numerous hybrid equations of state (EOS) that satisfy the observational data, in order to explore the possible existence of hybrid stars.
We adopt the two-model approach,
in which the Relativistic Mean Field (RMF) model, as in \cite{constrain_apply},
describes the hadron phase,
and the quiral symmetric SU(3) Nambu-Jona-Lasinio (NJL) model \cite{Klevansky:1992qe,Hatsuda:1994pi,Buballa:1998pr} with four-quark and eight-quark interaction terms \cite{renan_ww, renan_wr}
describes the quark phase. We also obtain hybrid EOS sets using the Mean Field Approximation of QCD (MFTQCD). {The MFTQCD model is} derived {from the QCD Lagrangian, considering a  gluon field decomposition in soft and 
hard momentum components. A mean field approximation is considered for the hard
gluons and its contribution stiffens the EOS allowing for high pressures at high energy densities. The soft gluon fields  originate condensates that  soften the EOS and give rise to a bag like term \cite{Fogaca:2010mf}.}

In 1961, before the formulation of QCD, Yochiro Nambu and Giovanni Jona-Lasinio proposed
the Nambu-Jona-Lasinio (NJL) model \cite{Nambu:1961tp, Nambu:1961fr}.
In the original idea, the Dirac particle mass was assumed to be produced
by the nucleon interaction, analogous to the superconductivity in BCS theory.
Later, NJL model was adapted to have quarks and gluons as degrees of freedom.
It became an effective theory of QCD, being a good approximation in the low energy limit.
In the NJL model, gluons are omitted and the quark interaction is assumed to be point-like.
Although color confinement property can not be described by NJL,
it reproduces all the QCD global symmetries and the spontaneus chiral symmetry breaking, see \cite{Buballa:2003qv} for a review.
The NJL model is suitable in the context of NS and
has been widely used to describe the deconfined phase in hybrid stars
\cite{Buballa:1998pr, Schertler:1999xn, Baldo:2002ju, Menezes:2003xa, Shovkovy:2003ce, Pagliara:2007ph, Bonanno:2011ch, renan_w_r, renan_wr, renan_ww}.

Some of the studies discussing hybrid  stars also apply a Bayesian inference to constrain the model parameters, either using a grid approach \cite{Alvarez-Castillo:2016oln,Pfaff:2021kse,Ayriyan:2021prr,Takatsy:2023xzf} or a Markov-Chain Monte Carlo
sampling process \cite{Xie:2020rwg}. Different models are used, a metamodeling description for the hadronic \cite{Pfaff:2021kse,Xie:2020rwg}, or a RMF model \cite{Alvarez-Castillo:2016oln} together with  a quark phase  described by the NJL model \cite{Alvarez-Castillo:2016oln,Pfaff:2021kse} or  a quark metamodel \cite{Xie:2020rwg}, in all cases describing the deconfinement phase transition by a Maxwell construction, or a parametrized smooth crossover between the hadronic phase described by a RMF EOS and a quark phase described within a constituent quark model  \cite{Takatsy:2023xzf}.

In a previous work \cite{constrain_apply},
Bayesian inference was used to explore the NSs EOS with a
RMF model.
However, it was assumed that NS were made entirely by hadron matter.
In this work, we explore the possibility of the existence of quark matter inside NS using microscopic models.
For the hadron phase, we choose two equations of state from \cite{constrain_apply} with extreme properties, in particular, a soft and a stiff EOS.
The quark phase is described by the Nambu-Jona-Lasinio (NJL) model and the MFTQCD model.
Bayesian inference is used to obtain large samples  of  EOS that satisfy
 the imposed constraints,
which include
the NICER x-ray observations from PSR J0030+0451 \cite{Riley:2019yda, Miller:2019cac}
and PSR J0740+6620 \cite{Riley:2021pdl, Miller:2021qha}
and a constraint in the phase transition density, which we consider should occur above saturation density. 
In \cite{Komoltsev2021jzg} it was found that
pQCD calculations can have an influence at intermediate densities, if thermodynamic relations and causality are taken into account. Therefore, we also impose these pQCD constraints in some of the generated sets
to analyze their effect on the results for NSs.

This article is structured as follows: Section \ref{chap:model} gives a brief overview of the NJL, MFTQCD and RMF models.  The Bayesian inference is described in Section \ref{chap:bayesian}. The results are presented and discussed in Section \ref{chap:results} while the  conclusions are drawn in Section \ref{chap:conclusion}.

\section{Model}\label{chap:model}
In this section, we present the quark and the hadron models separately.
To build hybrid equations, we assumed a first order quark-hadron phase transition
and applied the Maxwell construction.
Instead, a Gibbs construction could have been used, which considers charge neutrality globally rather than locally, as discussed in \cite{glendenning}. However, the Gibbs construction  proposed in \cite{glendenning} neglects finite size effects such as nuclear finite range and Coulomb effects. By including finite size effects in the description, it has been shown that the extension of the transition region depends on the surface tension, and when this property takes a large value, the Gibbs transition is close to a Maxwell construction \cite{Maruyama:2007ey}. Since the surface tension of quark clusters in hadronic matter is still unknown, a Maxwell construction valid in the limit of large surface tensions is applied.

\subsection{Quark phase}
We obtained sets considering the Nambu-Jona-Lasinio (NJL) model (Section \ref{sec:NJL})
and sets considering the Mean Field Theory of QCD (Section \ref{sec:mQCD}).
A brief review of both models is presented here.

\subsubsection{Nambu-Jona-Lasinio (NJL)} \label{sec:NJL}
In this work, we considered the SU(3) Nambu-Jona-Lasinio (NJL) Lagrangian being as
\begin{align}
  \mathcal{L}_\text{NJL}
  &= \bar{\psi}
    \left(
      i \slashed{\partial} -m +\mu \gamma^0
    \right) \psi \nonumber \\
  &\quad +\frac{G}{2}
    \left[
      \left( \bar{\psi} \lambda_a \psi \right)^2
      +\left( \bar{\psi} i \gamma^5 \lambda_a \psi \right)^2
    \right] \nonumber \\
  &\quad +\mathcal{L}_\text{'t Hooft} +\mathcal{L}_\text{I},
  \label{eq:Lnjl}
\end{align}
which $\mathcal{L}_\text{'t Hooft}$ is known as the 't Hooft term,
to simulate the U(1)$_A$ symmetry breaking
\begin{equation}
  \mathcal{L}_\text{'t Hooft}
  = 8 \kappa
    \left[
      \det \left( \bar{\psi} P_R \psi \right)
      +\det \left( \bar{\psi} P_L \psi \right)
    \right],
  \label{eq:LT}
\end{equation}
where the determinant is applied in the space of flavors.
However, these terms are not enough to reach the $2 M_\odot$.
For this reason, we add 4- and 8-quark interaction terms:
\begin{align}
  \mathcal{L}_\text{I}
  &= -G_\omega
    \left[
      \left( \bar{\psi} \gamma^\mu \lambda_0 \psi \right)^2
      +\left( \bar{\psi} \gamma^\mu \gamma_5 \lambda_0 \psi \right)^2
    \right] \nonumber \\
  &\quad -G_\rho \sum_{a=1}^8
    \left[
      \left( \bar{\psi} \gamma^\mu \lambda^a \psi \right)^2
      +\left( \bar{\psi} \gamma^\mu \gamma_5 \lambda^a \psi \right)^2
    \right] \nonumber \\
  &\quad -G_{\omega\omega}
    \left[
      \left( \bar{\psi} \gamma^\mu \lambda_0 \psi \right)^2
      +\left( \bar{\psi} \gamma^\mu \gamma_5 \lambda_0 \psi \right)^2
    \right]^2 \nonumber \\
  &\quad -G_{\sigma\omega} \sum_{a=0}^8
    \left[
      \left( \bar{\psi} \lambda_a \psi \right)^2
      + \left( \bar{\psi} i \gamma^5 \lambda_a \psi \right)^2
    \right] \nonumber \\
    &\qquad \times
    \left[
      \left( \bar{\psi} \gamma^\mu \lambda_0 \psi \right)^2
      + \left(
        \bar{\psi} \gamma^\mu \gamma_5 \lambda_0 \psi
      \right)^2
    \right] \nonumber \\
  &\quad -G_{\rho\omega} \sum_{a=1}^8
    \left[
      \left( \bar{\psi} \gamma^\mu \lambda_0 \psi \right)^2
      +\left(\bar{\psi} \gamma^\mu \gamma_5 \lambda_0 \psi \right)^2
    \right] \nonumber \\
    &\qquad \times
    \left[
      \left( \bar{\psi} \gamma^\mu \lambda_a \psi \right)^2
      +\left(\bar{\psi} \gamma^\mu \gamma_5 \lambda_a \psi \right)^2
    \right].
  \label{eq:LI}
\end{align}
Most of the interaction terms effects were studied in \cite{renan_w_r, renan_ww, renan_wr}.
As the NJL model is not renormalizable, we implemented the 3-momentum cutoff scheme.
After the mean field approximation, the grand canonical potential at $T = 0$ can be written as
\begin{align}
  \Omega
  &= \Omega_0
    +G
    \left(
      \sigma_u^2 + \sigma_d^2 + \sigma_s^2
    \right)
    +4 \kappa \sigma_u \sigma_d \sigma_s \nonumber \\
  &\quad 
    -\frac{2}{3} G_\omega (\rho_u + \rho_d + \rho_s)^2 \nonumber \\
  &\quad 
    -\frac{4}{3} G_\rho
      (\rho_u^2 +\rho_s^2 +\rho_s^2 -\rho_u \rho_d -\rho_u \rho_s -\rho_d \rho_s) \nonumber \\
  &\quad
    -\frac{4}{3} G_{\omega\omega} (\rho_u + \rho_d + \rho_s)^4 \nonumber \\
  &\quad
    -4 G_{\sigma\omega}
      \left( \sigma_u^2 + \sigma_d^2 + \sigma_s^2 \right)
      \left( \rho_u + \rho_d + \rho_s \right)^2
    \nonumber \\
  &\quad -\frac{8}{3} G_{\rho\omega}
    \left( \rho_u + \rho_d + \rho_s \right)^2 \nonumber \\
  &\qquad \times
    \left(
      \rho_u^2 + \rho_d^2 + \rho_s^2
      -\rho_u \rho_d -\rho_u \rho_s -\rho_d \rho_s
    \right) \nonumber \\
  &\quad -\frac{3}{\pi^2}
    \sum_{f=u,d,s} \int_{k_{F_f}}^\Lambda dp p^2 E_f
    -\frac{1}{\pi^2} \sum_{f=u,d,s} \tilde{\mu}_f k_{F_f},
  \label{eq:Omeganjl}
\end{align}
where $k_{F_f} = \sqrt{\tilde{\mu}_f^2 -\tilde{m}_f^2}$ is the Fermi momentum
and $\Omega_0$ is calculated to vanishs the potential in the vacuum.
The effective mass ($\tilde{m}$) and chemical potential ($\tilde{\mu}$) expressions
are given by the gap equations
\begin{align}
  \tilde{m}_i &= m_i + -2 G \sigma_i -2 \kappa \sigma_j \sigma_k \nonumber \\
    &\quad
    +\frac{8}{3} G_{\sigma\omega} (\sigma_i^2 +\sigma_j^2 +\sigma_k^2) \sigma_i \\
  \tilde{\mu}_i &= \mu_i - 
  \frac{4}{3} G_\omega (\rho_i +\rho_j +\rho_k)
    -\frac{4}{3} G_\rho (2 \rho_i -\rho_j -\rho_k) \nonumber \\
    &\quad
    -\frac{16}{9} G_{\omega\omega} (\rho_i +\rho_j +\rho_k)^3 \nonumber \\
    &\quad
    -\frac{8}{3} G_{\sigma\omega} (\sigma_i^2 +\sigma_j^2 +\sigma_k^2)
      (\rho_i +\rho_j +\rho_k) \nonumber \\
    &\quad
    -\frac{8}{9} G_{\rho\omega} (\rho_i +\rho_j +\rho_k) \nonumber \\
    &\quad
    \times (4 \rho_i^2 +\rho_j^2 +\rho_k^2 -\rho_i \rho_j -\rho_i \rho_k -4 \rho_j \rho_k),
\end{align}
for ${i \neq j \neq k} \in {u, d, s}$.
Imposing thermodynamic consistence \cite{buballa}, i.e., $d \Omega / d\sigma_f = 0$
and $d \Omega / d \rho_f = 0$, we obtain expressions for the quark condensate ($\sigma_i$)
and density ($\rho_i$). At zero temperature, we can write them as
\begin{align}
	\sigma_f &=
	   -\frac{3}{\pi^2}
	   \int_{k_{F_f}}^\Lambda dp p^2 \frac{\tilde{m}_f}{E_f}, \\
	\rho_f &= \frac{1}{\pi^2} k_{F_f}^3.
\end{align}

The parameters are set in order to reproduce
the experimental values of mass and decay constant
from $\pi^\pm$, $K^\pm$, $\eta$ and $\eta'$ mesons
obtained in \cite{Olive_2014}.
Their values are given in Table \ref{tab:fixed_njl}.
In addition, we also use the electron mass being $m_e = 0.510 \text{ MeV}$.

\begin{table}[]
\setlength{\tabcolsep}{7.pt}
\renewcommand{\arraystretch}{1.2}
\begin{tabular}{ccccc}
\hline \hline 
$\Lambda$ (MeV) & $m_{u,d}$ (MeV) & $m_s$ (MeV) & $G \Lambda^2$ & $\kappa \Lambda^5$ \\ \hline
623.58 & 5.70 & 136.60 & 3.34 & -13.67 \\
\hline
\end{tabular}
\caption{Fixed parameters of NJL model.
  $\Lambda$ is the cutoff, $m_{u,d,s}$ are the quarks up, down and strange masses,
  $G$ and $\kappa$ are coupling constants.
  The values of these parameters are obtained in such a way that
  they reproduce the experimental masses and decay constants
  of mesons \cite{Olive_2014}.}
\label{tab:fixed_njl}
\end{table}

As the NJL is normalized in the vacuum, we add a constant $B$ in the pressure equation,  $P \to P +B$, similar to the MIT bag model. {This term was also included in \cite{Pagliara:2007ph,Bonanno:2011ch}, in the first study to allow the restoration of chiral symmetry and the deconfinement phase transition to occur at the same baryon density, and in the second to control the deconfinement phase transition. In the present study we include the extra $B$ term with the same proposal as in \cite{Bonanno:2011ch}.}
In this way, we have 6 parameters to be determined by Bayesian analysis:
the adimensional coupling constants (defined as $\xi_i = 2 G_i / G$ and $\xi_{ii} = 16 G_{ii} / G^4$)
$\xi_\omega$, $\xi_\rho$, $\xi_{\omega\omega}$, $\xi_{\omega\sigma}$,
$\xi_{\omega\rho}$ and the constant $B$.

\subsubsection{Mean Field Theory of QCD (MFTQCD)} \label{sec:mQCD}

In this model, we start with the QCD lagrangian
and we assume we can decompose the gluon field in
low and high momentum components (soft and hard gluons) \cite{Fogaca:2010mf, Franzon:2012in, Albino:2021zml}
\begin{equation}
    \tilde{G}^{a\mu} (k) = \tilde{A}^{a\mu} (k) + \tilde{\alpha}^{a\mu} (k),
\end{equation}
where $\tilde{G}$ is the gluon field in the momentum space
and we defined $A$ and $\alpha$ as the soft and hard gluon, respectively.
We approximate the momentum of the soft gluons to be zero.
This implies that their field is constant.
We replace the fields with their expected value in the vacuum
\cite{PhysRevD.34.1591, PhysRevD.71.074007}
\begin{align}
    \left< A^{a\mu} A^{b\nu} \right>
        &= -\frac{\delta^{ab}}{8} \frac{g^{\mu\nu}}{4} \mu_0^2, \\
    \left< A^{a\mu} A^{b\nu} A^{c\rho} A^{d\eta} \right>
        &= \frac{\phi_0^4}{(32)(34)}
        \left[
            g_{\mu\nu} g^{\rho\eta} \delta^{ab} \delta^{cd} 
        \right. \nonumber \\
        &\quad \left. 
            +g_\mu^\rho g_\nu^\eta \delta^{ac} \delta^{bd} 
            +g_\mu^\eta g_\nu^\rho \delta^{ad} \delta^{bc} 
        \right],
\end{align}
where $\phi_0$ and $\mu_0$ are the energy scale that will be determined.
Next, we consider the hard gluons as classical field,
due to its large occupation numbers at all energy levels
\cite{Serot:1984ey, Tezuka:1987tn}
\begin{equation}
    \alpha_\mu^a \to \left< \alpha_\mu^a \right> = \alpha_0^a \delta_{\mu 0},
\end{equation}
where $\alpha_0^a$ is constant.
After some calculation, we obtain the MFTQCD lagrangian
\begin{align}
	\mathcal{L}_\text{MFTQCD} &=
        -B
        +\frac{m_G^2}{2} \alpha_0^a \alpha_0^a
        \nonumber \\ &\quad
		+\sum_{q=1}^{N_f} \bar{\psi}_i^q
		\left( 
			i \delta_{ij} \gamma^\mu \partial_\mu
			+ g \gamma^0 T_{ij}^a \alpha_0^1
			-\delta_{ij} m_q
		\right) \psi_j^q, 
\end{align}
where we defined
\begin{align}
    m_G^2 &= \frac{9}{32} g^2 \mu_0^2, \\
    B &= \frac{9}{4(34)} g^2 \phi_0^4 = \left< \frac{1}{4} F^{a\mu\nu} F^a_{\mu\nu} \right>.
\end{align}
The 2-dimensional condensate contributes to the mass of the hard gluon ($m_G$),
while the 4-dimensional one acts as a constant analogous to the MIT bag model.
After obtaining the equations of motion and using the energy-momentum tensor,
we obtain the equations of state below:
\begin{align}
	P &=  \frac{27}{2} \xi^2 \rho_B^2 -B +P_F, \\
	\epsilon &= \frac{27}{2} \xi^2 \rho_B^2 +B +\epsilon_F,
    \label{eq:mqcd}
\end{align}
where $P_F$ and $\epsilon_F$ are the pressure and energy density
of a non-interacting Fermi gas of quarks and electrons and $\xi = g/m_G$.
The deduction of the EOS can be found in \cite{Fogaca:2010mf}.
We considered the masses values as $m_u = 5 \text{ MeV}$, $m_d = 7 \text{ MeV}$
and $m_s = 100 \text{ MeV}$.
$\xi$ and $B$ are determined from the Bayesian analysis.

\subsection{Hadron phase}
In \cite{constrain_apply}, equations of state of pure nuclear matter were obtained
from a relativistic mean field model with non-linear meson terms,
both self-interactions and mixed terms.
In this model, the nuclear interactions are described by the exchange
of the $\sigma$ (scalar-isoscalar), $\omega$ (vector-isoscalar) and $\rho$ (vector-isovector) mesons.
The Lagrangian is given by
\begin{equation}
    \mathcal{L} = \mathcal{L}_N +\mathcal{L}_M +\mathcal{L}_{NL},
\end{equation}
where
\begin{align}
    \mathcal{L}_N &= \bar{\Psi}
        \left[
            \gamma^\mu \left(
                i \partial_\mu -g_\omega \omega_\mu -g_\rho \boldsymbol{t} \cdot \boldsymbol{\rho}_\mu
            \right)
            \right. \nonumber \\
            &\quad \left.
            -(m -g_\sigma \phi)
        \right] \Psi,  \\
    \mathcal{L}_M &= \frac{1}{2}
        \left[
            \partial_\mu \phi \partial^\mu \phi
            -m_\sigma^2 \phi^2
        \right] \nonumber \\
        &\quad
        -\frac{1}{4} F_{\mu\nu}^{(\omega)} F^{(\omega)\mu\nu}
        +\frac{1}{2} m_\omega^2 \omega_\mu \omega^\mu \nonumber \\
        &\quad
        -\frac{1}{4} \boldsymbol{F}_{\mu\nu}^{(\rho)} \cdot \boldsymbol{F}^{(\rho)\mu\nu}
        +\frac{1}{2} m_\rho^2 \boldsymbol{\rho}_\mu \cdot \boldsymbol{\rho}^\mu, \\
    \mathcal{L}_{NL} &=
        -\frac{1}{3} b m g_\sigma^3 (\sigma)^3
        -\frac{1}{4} c g_\sigma^4 (\sigma)^4 
        +\frac{\xi}{4!} g_\omega^4 (\omega_\mu \omega^\mu)^2 \nonumber \\
        &\quad
        +\Lambda_\omega g_\rho^2 \boldsymbol{\rho}_\mu \cdot \boldsymbol{\rho}^\mu
        g_\omega^2 \omega_\mu \omega^\mu,
\end{align}
where the Dirac spinor $\Psi$ represents the nucleon doublet (neutron and proton)
with a bare mass $m$;
$g_\sigma$ ($m_\sigma$), $g_\omega$ ($m_\omega$) and $g_\rho$ ($m_\rho$)
are the coupling constants (masses) of the $\sigma$, $\omega$ and $\rho$ mesons, respectively; 
$b$, $c$, $\xi$ and $\Lambda_\omega$  are the couplings that determine the strength of the non-linear terms interaction.
All coupling constants were obtained through a Bayesian inference approach.
This method was applied so that the equations should satisfy the
chiral EFT constraints for pure neutron matter, properties of nuclear matter and neutron star observational data. 
In this work, we will use two equations of state (EOS) from \cite{constrain_apply}:
the BMPF 220, a soft EOS, and the BMPF 260, a stiff EOS.
Nuclear matter and NSs properties of these equations are shown in Table
\ref{tab:properties_rmf}.
With this choice we  analyze the extreme hadronic cases, and consider they will define for a given mass the low and high radius extremes.

In the present study, hadronic matter is described without hyperons.  The presence of hyperons softens the EOS, and therefore the transition to deconfinement matter occurs at a higher density, see for example \cite{JiaJieLi2020}, favoring a smaller quark core.  However, if the quark EOS is hard enough, two solar masses are still reached, see for example \cite{JiaJieLi2020}, and therefore the inclusion of hyperons should also be investigated.

\begin{table}[]
 \setlength{\tabcolsep}{12.pt}
\renewcommand{\arraystretch}{1.2}
\begin{tabular}{cccc}
\hline \hline 
Quantity           & Units     & BMPF220 & BMPF260 \\ \hline
$\rho_0$           & fm$^{-3}$ & 0.146  & 0.149    \\
$m^\ast$           & …         & 0.74   & 0.61     \\
$\epsilon_0$       & MeV       & -15.93 & -16.02   \\
$K_0$              & MeV       & 288    & 244      \\
$J_{\text{sym},0}$ & MeV       & 31.74  & 29.06    \\
$L_{\text{sym},0}$ & MeV       & 45     & 56       \\
$K_{\text{sym},0}$ & MeV       & -159   & 2        \\
$M_{\text{max}}$   & M$_\odot$ & 2.200  & 2.592    \\
$R_{\text{max}}$   & km        & 11.35  & 12.64    \\
$R_{1.4}$          & km        & 12.94  & 13.51    \\
$\Lambda_{1.4}$    &           & 548    & 828      \\ \hline
\end{tabular}
\caption{Nuclear matter properties and NSs properties of the hadronic EOS.
\label{tab:properties_rmf}}
\end{table}

\section{Bayesian Approach}\label{chap:bayesian}
The Bayesian method is a robust technique that enables us to sample an ensemble of EOS, all of which meet the specified constraints. To apply this, we have to define the prior, likelihood, and sampler.

\textit{Prior:-} In Bayesian inference, a prior distribution represents our initial assumptions about a parameter before any data is observed. This prior is then revised with new data through the likelihood of the observed data to create the posterior distribution. In our calculation, we have employed uniform priors for all the parameters of the model.  The minimum and maximum values of the uniform distribution are shown in Table \ref{tab:prior_njl}.

\begin{table*}[]
 \setlength{\tabcolsep}{12.pt}
\renewcommand{\arraystretch}{1.2}
\begin{tabular}{ccccccccc}
\hline \hline 
\multicolumn{4}{c}{NJL}                          &  & \multicolumn{4}{c}{MFTQCD}                \\ 
Parameters           & Units         & min & max &  & Parameters & Units         & min & max    \\ \hline
$\xi_\omega$         & …             & 0   & 0.6 &  & $\xi$      & MeV$^{-1}$    & 0   & 0.0018 \\
$\xi_\rho$           & …             & 0   & 1   &  & $B$        & MeV.fm$^{-3}$ & 50  & 180    \\
$\xi_{\omega\omega}$ & …             & 0   & 40  &  &            &               &     &        \\
$\xi_{\sigma\omega}$ & …             & 0   & 8   &  &            &               &     &        \\
$\xi_{\rho\omega}$   & …             & 0   & 50  &  &            &               &     &        \\
$B$                  & MeV.fm$^{-3}$ & 0   & 50  &  &            &               &     &        \\ \hline
\end{tabular}
\caption{The lowest and highest values for the uniform distribution prior used for NJL and MFTQCD in this study. \label{tab:prior_njl}}
\end{table*}

\textit{Likelihood:-} In Bayesian inference, the likelihood quantifies the probability of the observed data for particular parameter values, acting as a link between prior beliefs and posterior knowledge.

\begin{enumerate}[(i)]
\item \textbf{X-ray observation(NICER):} X-ray observations give the mass and radius measurements of NS. Therefore, the corresponding evidence takes the following form,
\begin{align}
    P(d_{\rm X-ray}|\mathrm{EOS}) = \int^{M_{\mathrm{max} }}_{M_{\mathrm{min} }} dm P(m|\mathrm{EOS}) \nonumber \\ \times
    P(d_{\rm X-ray} | m, R (m, \mathrm{EOS})) \nonumber \\
    = \mathcal{L}^{\rm NICER}
\end{align}

\begin{equation}
    P(m|\rm{EOS}) = \left\{ \begin{matrix} \frac{1}{M_\mathrm{max} - M_\mathrm{min}} & \text{ if  } M_\mathrm{min} \leq m \leq M_\mathrm{max}\ , \\ 0 & \text{otherwise.} & \end{matrix} \right .
\end{equation}
{Here, $M_{\rm min}$ is 1 $M_\odot$, and $M_{\rm max}$ represents the maximum mass of a NS for the given EOS \cite{Landry:2020vaw}.}

\item \textbf{Phase transition:} We constraint the phase transition value through a 
super-Gaussian centered at $\rho_\text{trans} = 0.275 \text{ fm}^{-3}$,
with a standard deviation of $\sigma = 0.08 \text{ fm}^{-3}$ and $p = 5$.
Therefore, the likelihood is written as
\begin{equation}
  \mathcal{L}^{\rm PhT} =
    \text{exp} \left[
        - \left(
            \frac{(\rho_\text{trans} -\rho_B)^2}{2 \sigma^2}
        \right)^p
        \right].
\end{equation}
The density range was chosen such that the occurrence of the transition is possible in medium mass stars with $M\geq 1.4 M_\odot$. If the transition occurs at a too large density the quark core is too small or does not exist inside a NS. For the lower limit, saturation density was considered. Since we are dealing with extremely asymmetric matter, with a proton fraction below 10\% it is reasonable to suppose that a transition to quark matter could be favored energetically, see for instance \cite{Cavagnoli:2010yb} where  a phase transition to deconfined matter below twice saturation density has been obtained for some models.  This extreme lower limit  has  also been used in the literature to explore the properties of hybrid stars, see for instance \cite{JiaJieLi2020,Raithel:2022efm}.

\item \textbf{pQCD EOS:}
In \cite{Komoltsev2021jzg}, the authors found that for the neutron stars' EOS to be in 
agreement with the pQCD calculations, they must be within a limited region in
$P \times \epsilon \times \rho_B$ space. 
To apply this restriction,
we have calculated the area in $P \times \epsilon$
allowed by pQCD at $\rho_B = 7 \rho_0$ for the renormalization scale $X = 4$.
Thus, we define the likelihood as
\begin{equation}
    \mathcal{L}^{\rm pQCD} =
    \begin{cases}
        1, (P(7\rho_0),\epsilon(7\rho_0))
        \text{ inside pQCD region}; \\
        0, (P(7\rho_0),\epsilon(7\rho_0))
        \text{ \textbf{not} inside},
    \end{cases}
\end{equation}
where $(P(7\rho_0),\epsilon(7\rho_0))$ is the pressure and energy density
at $7 \rho_0$.

\end{enumerate}

The final likelihood is then given by
\begin{equation}
    \mathcal{L} = 
    \mathcal{L}^{\rm NICER I}
    \mathcal{L}^{\rm NICER II}
    \mathcal{L}^{\rm PhT}
    \mathcal{L}^{\rm pQCD}\,.
    \label{eq:finllhd}
\end{equation}

NICER I and NICER II  correspond to the mass-radius measurements of PSR J0030+0451 and PSR J0740+6620, respectively.

\textit{Sampler:-} We utilized the nested sampling method, specifically the PyMultinest \cite{pymultinest, pymultinest2} sampler, as part of the Bayesian Inference Library (BILBY) \cite{bilby}.

\section{Results}\label{chap:results}
In this section we present and discuss the properties of hybrid stars, considering different models to describe the total EOS. For the confined phase, we choose two representative hadronic EOS, one soft and one hard, but still compatible with the constraints imposed on the EOS. This choice allows us to estimate the effect of the hadronic EOS on the total EOS. For the deconfined phase, we consider the two different models, NJL and MFTQCD, introduced in Sections \ref{sec:NJL} and \ref{sec:mQCD}, respectively. This choice allows us to estimate the effects of the choice of a particular quark model. In order to understand the effect of imposing the thermodynamic and casuality constraints from the pQCD EOS, we have also {performed the inference analysis}
with and without this constraint in the following.
Note that the NJL model cannot be extended up to a density as large as the one where the pQCD EOS was calculated,
but following \cite{Komoltsev2021jzg} we can estimate the effects of this constraint on the hybrid EOS.
\begin{figure}[!ht]
  \centering
  \includegraphics[width=.95\linewidth]{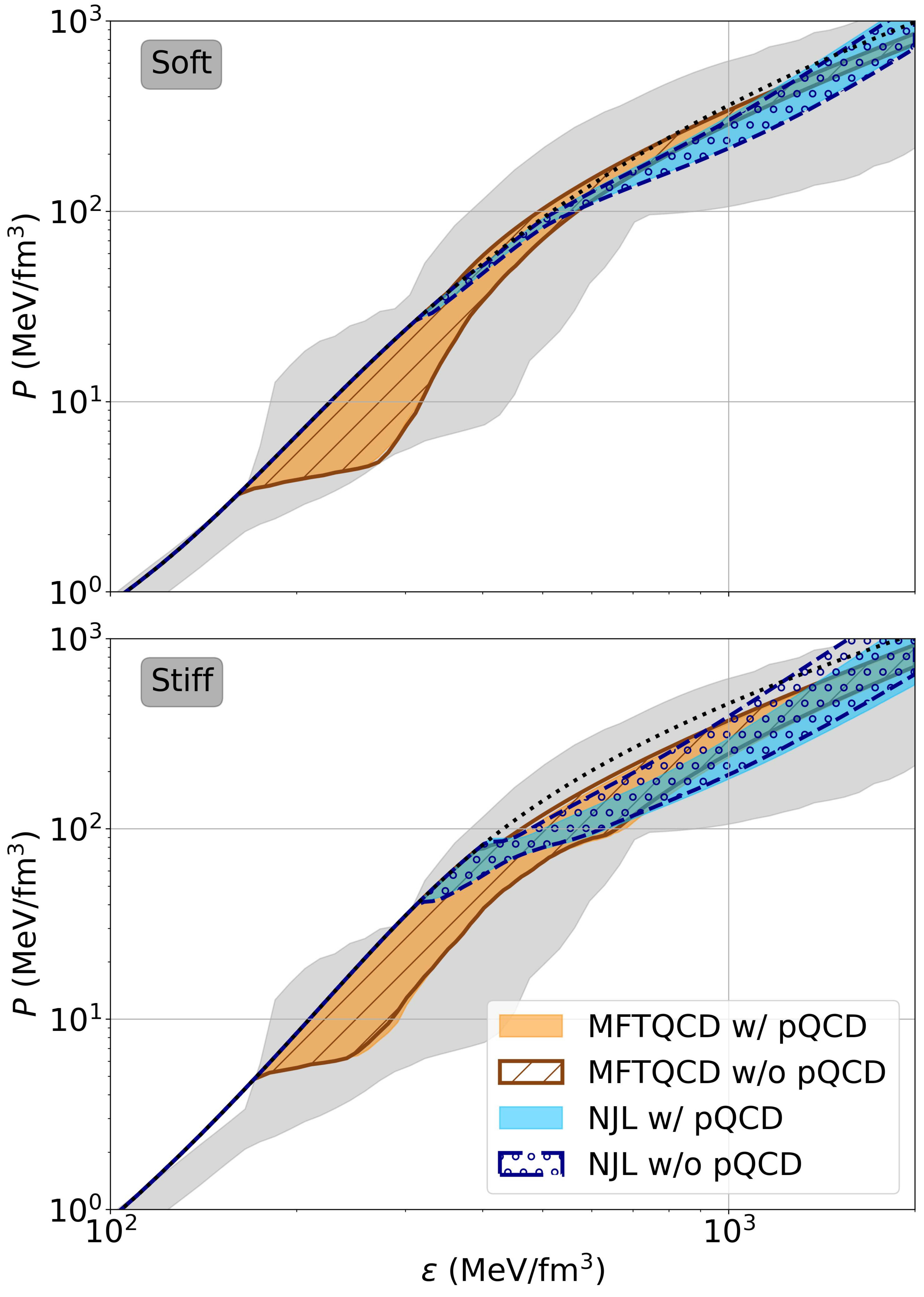}
  \caption{Pressure versus energy density for the soft EOS (left panel) and stiff EOS (right panel).
    Gray plots show the results from \cite{Annala:2019puf}.}
  \label{fig:graph_EoS}
\end{figure}

Taking into account the above choices, we have generated, within a Bayesian inference approach for the quark model parameters, eight different EOS sets with $\sim 8 000$ (NJL) and $\sim 2 000$ (MFTQCD) EOS each combining the following options:
i) two hadron EOS (soft and hard), ii) two quark models (NJL and MFTQCD), iii) with or without imposing the
pQCD constraint. These eight sets are discussed next.

In Fig. \ref{fig:graph_EoS}, the EOS distributions, i.e. the pressure versus the energy density, are shown at a  90\%  confidence level (CL) for the two hadronic models, the soft (stiff) on the left (right) panel. In both panels the hadronic EOS is represented by a  dotted line. The sets including (not including) the pQCD constraint are represented by colored homogeneous bands (patterns)
and the sets with the NJL (MFTQCD) model are represented by the colors blue/dark blue (orange/dark orange).
The gray band represents the results from \cite{Annala:2019puf}. This band spans the entire phase space generated by an agnostic description of the EOS and imposing constraints on pure neutron matter from $\chi$EFT calculations, the pQCD constraints at very high densities and  NS observables. 
Some conclusions are in order: i) our results are compatible with those obtained in \cite{Annala:2019puf} (shown in gray): all hybrid EOS fall within this range; ii) the MFTQCD model allows a deconfinement phase transition at much lower baryon densities compared to the NJL model
(phase transition distribution can be seen in Fig. \ref{fig:likelihood} in Appendix \ref{section:appendixA}).
Although, the constraints permitted a phase transition at a baryon density of the order of the saturation density or above, the transition  does not occur below $\rho_B\sim 0.25$~fm$^{-3}$ within NJL, while the MFTQCD model admits transitions at saturation density. 

\begin{figure}[!t]
  \centering
  \includegraphics[width=0.95\linewidth]{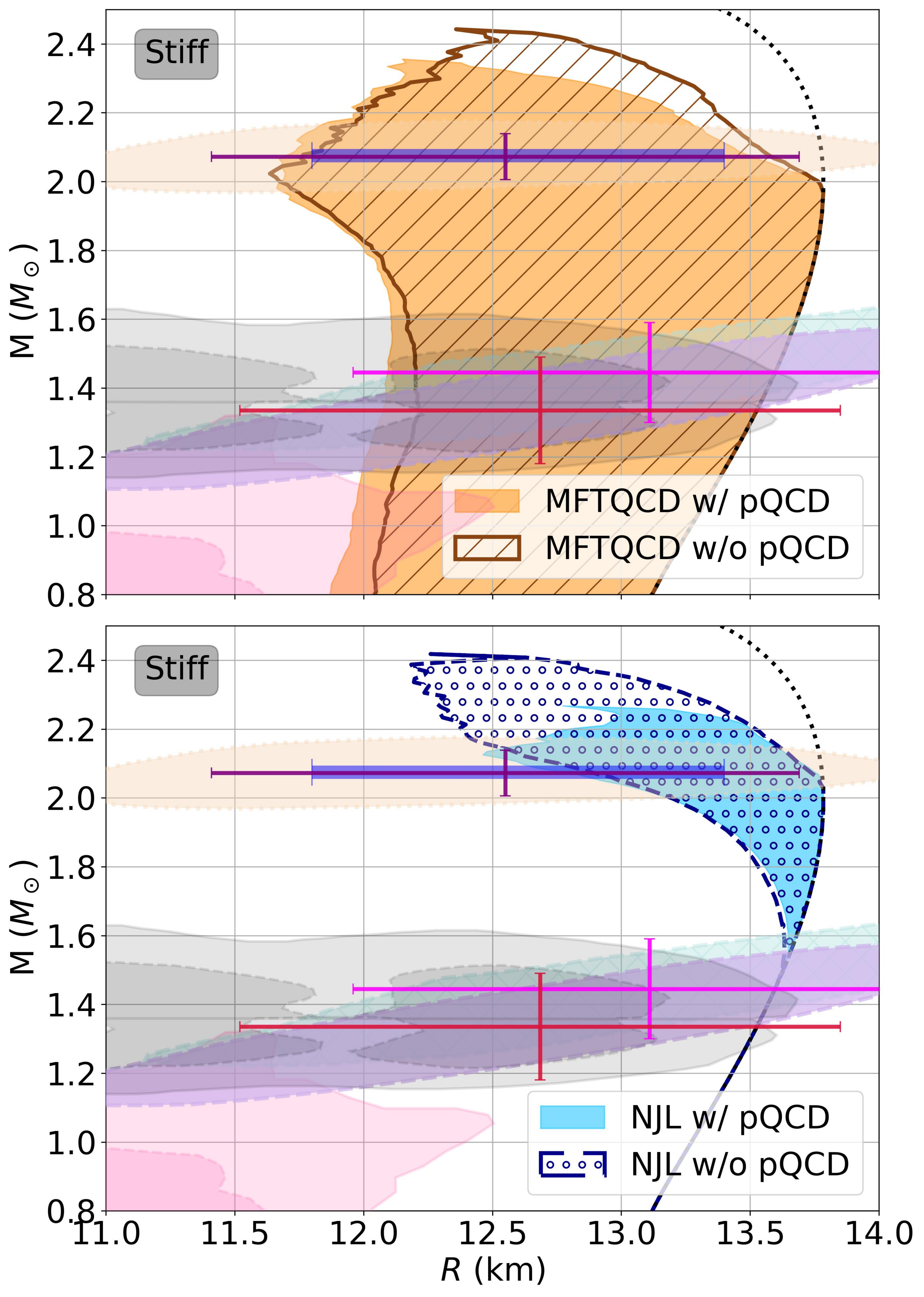}
  \caption{Mass-radius diagram of the 90\% of CL stiff sets: MFTQCD (left) and NJL (right).
  Solid bands represent sets were obtained applying the pQCD constraint.
  Observational data is show as:
  PSR J0030+0451 (cyan and purple) \cite{Riley:2019yda, Miller:2019cac},
  PSR J0740+6620 (yellow) \cite{Riley:2021pdl, Miller:2021qha},
  HESS J1731-347 (magenta) \cite{hess}, and GW170817 (light and dark gray) \cite{LIGOScientific:2018cki}.
  }
  \label{fig:graph_TOV}
\end{figure}

\begin{figure}[!t]
  \centering
  \includegraphics[width=.95\linewidth]{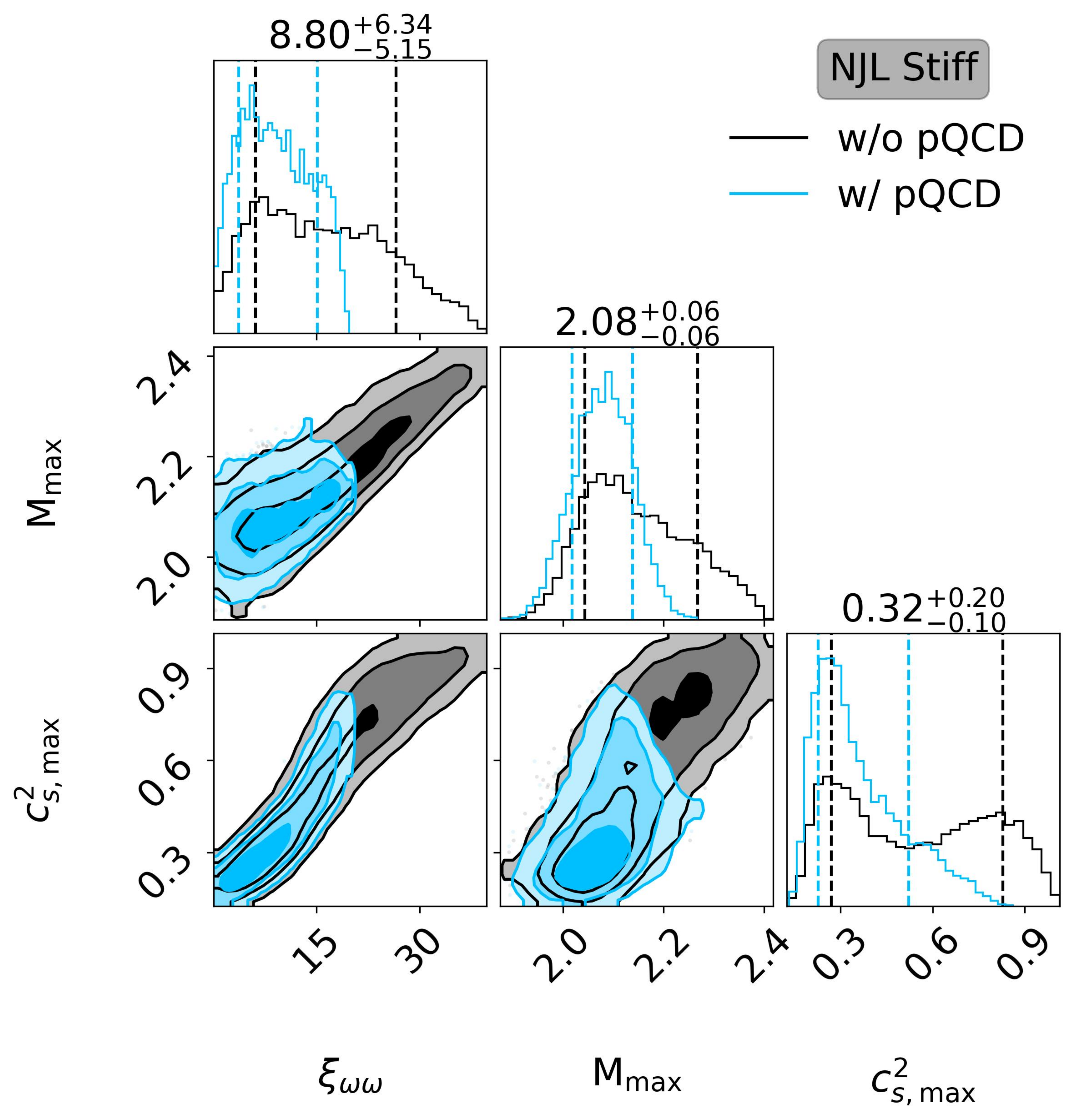}
  \caption{Corner plots of $\xi_{\omega\omega}$, the maximum mass ($M_\text{max}$) and
  the speed of sound at $M_\text{max}$ ($c^2_\text{s,max}$)
  for the NJL stiff sets.
  Set with (without) pQCD constraint is shown in blue (black).}
  \label{fig:corner_poster}
\end{figure}

\begin{figure*}[t]
  \centering
    \includegraphics[width=.95\linewidth]{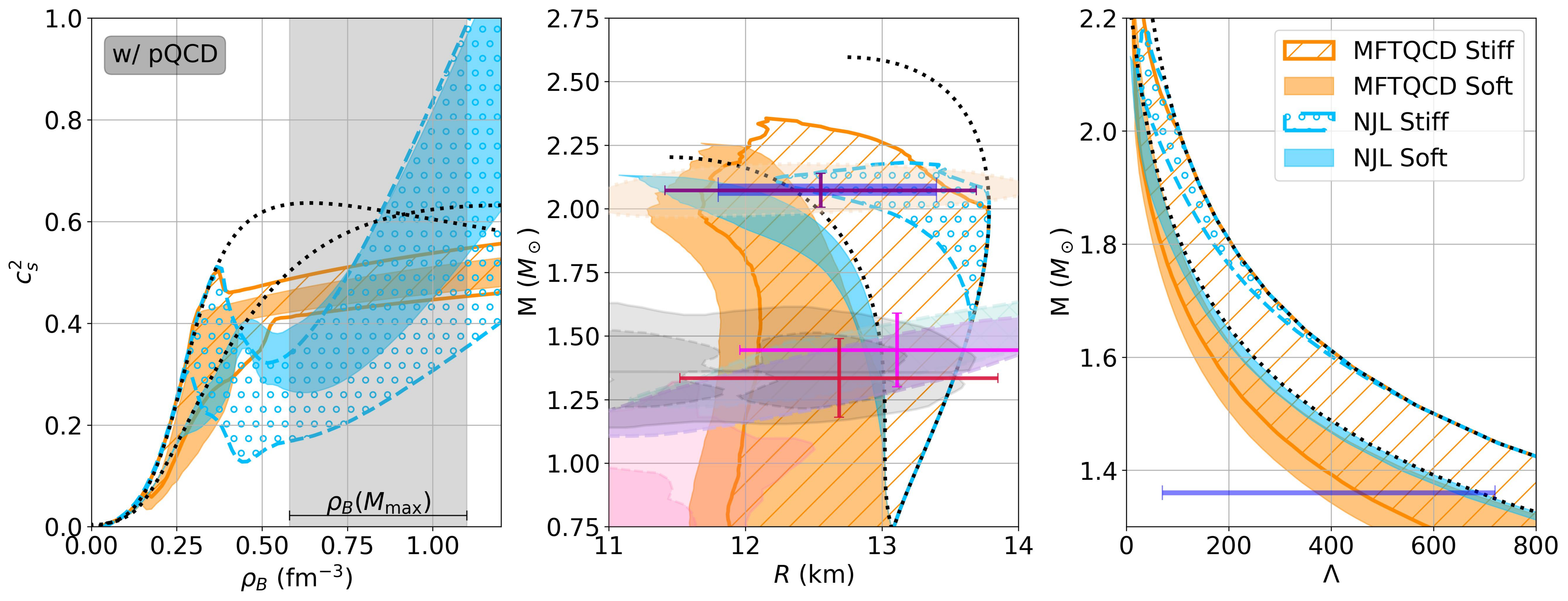}
  \caption{Comparison between models with pQCD constraint.
  It is shown, from the left to the right:
  the speed of sound as a function of the baryon density,
  the mass-radius diagram and
  the mass by the tidal deformability. The middle panel with the mass-radius curves also includes observational data as specified in Fig. \ref{fig:graph_TOV}.
  }
  \label{fig:all_pQCD}
\end{figure*}

\begin{figure}[!ht]
  \centering
  \includegraphics[width=.95\linewidth]{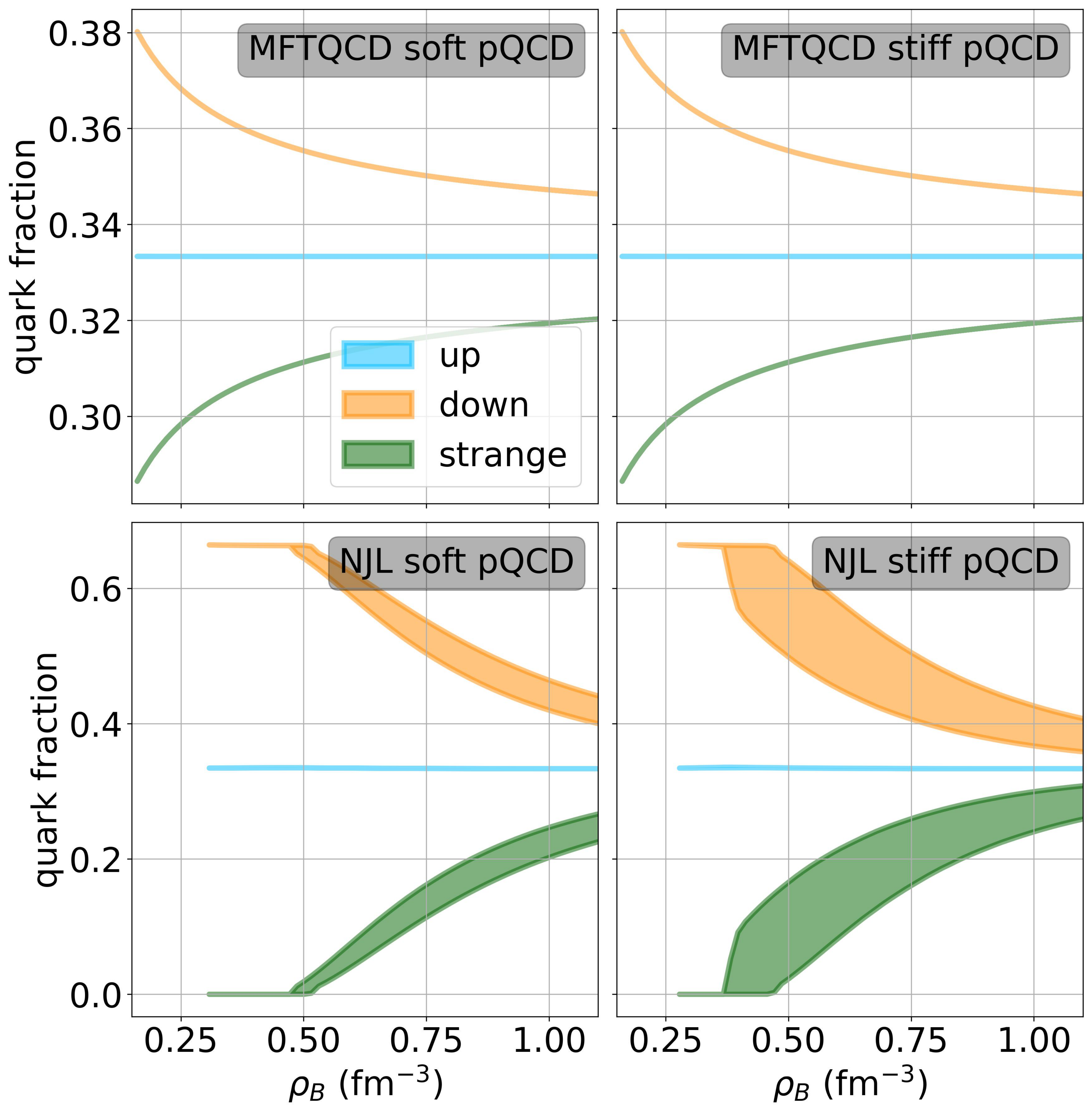}
  \caption{90\% CL of  the $u,\, d,\,$ and $s$-quark fraction, for MFTQCD (top) and NJL (bottom) and soft  (right) and stiff EOS (right). The bands at the low density limit define the region where the deconfinement occurs for the different set EOS, and result from the jump from zero to a finite value of the order of 1/3 for $u$-quarks and  $\lesssim2/3$ for the $d$-quark.}
  \label{fig:graph_frac}
\end{figure}

Solving the Tolman-Oppenheimer-Volkov equations \cite{Tolman:1939jz,Oppenheimer:1939ne},
we obtain the mass-radius relation of  static spherical and homogeneous neutron stars. 
Figure \ref{fig:graph_TOV} shows the mass-radius diagram  obtained with the sets  built from the stiff hadronic EOS.
The same color codes from Fig. \ref{fig:graph_EoS} are used.
In the mass-radius diagrams, we also show the 68\% confidence zone
of the 2D posterior distribution of
PSR J0030+0451 (cyan and magenta) \cite{Riley:2019yda, Miller:2019cac} and PSR J0740+6620 (yellow) \cite{Riley:2021pdl, Miller:2021qha}  NICER data, and the  HESS J1731-347 (magenta) \cite{hess}.
The GW170817 data from the LIGO-Virgo collaboration is represented by the
solid (dashed) gray area with 90\% (50\%) CL \cite{LIGOScientific:2018cki}.

 The pQCD constraint reduces the maximum mass in
both NJL and MFTQCD models.
However, this reduction occurs with more intensity in the stiff NJL model. 
This effect can be explained by the corner plot given in Fig. \ref{fig:corner_poster}.
In this figure, we show the relation between $\xi_{\omega\omega}$,
the maximum mass ($M_\text{max}$) and the speed of sound at the maximum mass ($c^2_{s,\text{max}}$)
for the NJL stiff set with (without) the pQCD constraint in blue (black).
We identify  a strong correlation between these terms.
Thus, $\xi_{\omega\omega}$ is the variable responsible for generating high values of maximum mass
at the cost of considerably increasing the values of the speed of sound.
However,  the pQCD constraint is related to causality \cite{Komoltsev2021jzg},
allowing only smaller $\xi_{\omega\omega}$.
We do not have this behavior in the soft sets,
as the $\xi_{\omega\omega}$ values are  smaller (see Figs. \ref{fig:corner_parameters_njl} and \ref{fig:corner_parameters_mqcd} in 
Appendix \ref{section:appendixA} for a corner plot with the model parameters).

The pQCD constraint has also a direct effect on the speed of sound. When this constraint is imposed, the NJL $c_s^2$ PDF shows a maximum  below $c_s^2=0.3$, and the medium value in the centre of the maximum mass configuration is 0.32, although it can reach values as large as $\sim 0.7$. On the other hand, for the soft hadron EOS a medium value $\sim 0.6$ is obtained, with the possibility that a value as large as $\sim 0.83$ is reached in the centre of the most massive stars (see Table \ref{tab:all}). In Table \ref{tab:all}, we give several NS properties (radius, transition density to deconfined quark matter, central pressure, energy density, baryon density and speed of sound squared) for stars with a mass equal to 1.4, 1.6, 1.8, 2.0 $M_\odot$ and maximum mass. 

None of the terms of MFTQCD model affect the speed of sound in a similar way as the NJL term $\xi_{\omega\omega}$. It is the non-linearity of this term in the NJL model involving the baryon density that causes the stiffening  of the EOS.  MFTQCD EOS predict  for  $c_s^2$ in the centre of the most massive stars  values of the order of 0.5, although they can decrease to $\sim 0.35$ in the case of the stiff EOS.

\begin{figure}[!ht]
  \centering
  \includegraphics[width=.95\linewidth]{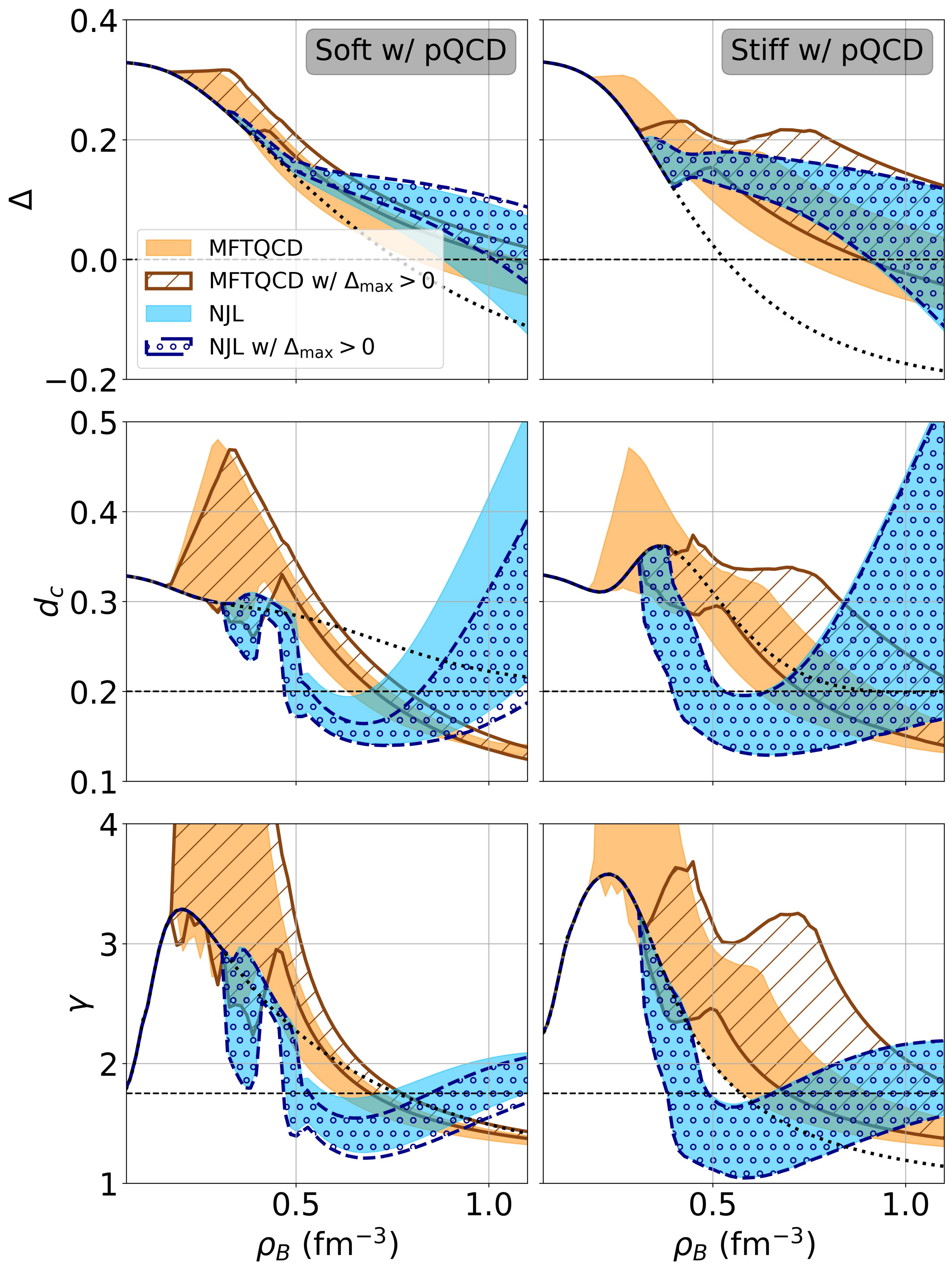}
  \caption{The top graphs show the $d_c$,
	and the graphs below show the speed of sound by baryonic density.
	In first column, we have model with the soft equation
	and in second column, with the stiff equation.}
  \label{fig:graph_trace}
\end{figure}

The information on the neutron star properties is completed  with Fig. \ref{fig:all_pQCD} which shows some properties of all the four sets including the pQCD constraint, soft and stiff hadron EOS with the NJL and MFTQCD  quark models: the speed of sound squared (left), the mass-radius relations (middle) and the tidal deformability as a function of the mass (right).
The NJL and the MFTQCD sets are shown in blue and orange, respectively.
Sets with the soft (stiff) equations are shown in solid (patterned) bands.
Observational data has the same code color introduced in Fig. \ref{fig:graph_TOV}.
In the tidal deformability versus  mass  graph (right panel), the GW170817 data from \cite{LIGOScientific:2018cki} is represented by the purple line.

Some comments are in order:
from the Fig. \ref{fig:graph_TOV} and the Table \ref{tab:all}, it is seen that both NJL and MFTQCD predict similar star radii for masses of the order of $2\, M_\odot$. However, radii below 12 km for medium mass stars can only be obtained with the MFTQCD model. While within the soft EOS, both quark models predict for the radius of the maximum mass configuration a value just above 11 km, within the stiff hadron EOS, the NJL predicts 13 km, more than 1 km larger than the MFTQCD model; ii) The pQCD constraints  imposes a maximum mass $\lesssim 2.3 \, M_\odot$ for MFTQCD and $\lesssim 2.1 \, M_\odot$ for NJL.  These constraints may lower  the maximum mass on about 0.2$M_\odot$ for NJL and 0.06$M_\odot$ for MFTQCD for the sets built with the stiff  hadronic EOS and has almost no effect for the soft EOS. See  Table \ref{tab:all} where some NS properties are given for all the sets built in the present work;
iii) 
All sets are compatible with the NICER observational data.
However, only MFTQCD can partially describe the HESS observation \cite{hess};
iv)
 The NJL stiff set is not able to describe the tidal deformability obtained from the GW170817 data, mainly because the stiff hadronic EOS does not satisfy this constraint. Moreover, the MFTQCD stiff is only partially compatible with it; v) for the NJL EOS, the deconfiment phase transition occurs just above 2 $\rho_0$ and does not vary much within the complete set. Within the MFTQCD, the deconfinement phase transition occurs mostly at a lower density, above the saturation density but below twice the saturation density; vi) for the NJL EOS, the speed of sound squared suffers a strong decrease at deconfinement, followed by a steady increase, reaching a value above 0.9 under some conditions. For the MFTQCD EOS, the increase in the speed of sound that follows the decrease at the phase transition is weak.

In Fig. \ref{fig:graph_frac}, the fraction of the three quark flavors is represented as a function of the density. For NJL,  the quark deconfinement occurs before the onset of the $s$-quark at $\sim 0.4$ (0.5) fm$^{-3}$  for the stiff (soft) EOS.  The deep in the speed of sound panel for the NJL model after deconfinement results from the $s$-quark onset. The large baryon density at which is occurs is due to the fact that the chiral restoration of  $s$-quark mass occurs at large densities. In the MFTQCD model, quarks are already in the chiral restored phase and it is the larger current mass of the $s$-quark the cause of a smaller $s$-quark fraction.

In the following we discuss some more properties of the EOS sets developed in the present study.
In \cite{Annala:2019puf,Fujimoto:2022ohj,Annala:2023cwx}, model independent equations were used to probe quark matter inside NSs.
The authors discussed the high density behavior of the EOS 
based on the polytropic index $\gamma=\frac{d\ln P}{d\ln \varepsilon}$,  on the normalized matter trace anomaly $\Delta=\frac{1}{3}-\frac{P}{\varepsilon}$ and some quantities derived from the trace anomaly. In \cite{Annala:2023cwx}, the authors have conjectured that the  matter part of the trace anomaly could be positive definite and discuss the implication on the mass-radius curves of NS.
In \cite{Annala:2023cwx}, the authors propose that $d_c = \sqrt{\Delta^2 + \Delta'^2} < 0.2$
(where $\Delta'$ is $\Delta$'s logarithmic derivative with respect to the energy density $\varepsilon$)
could indicate a deconfined matter behavior.
In this work, by using the microscopic models that define the different datasets, and since  the composition of matter is know, we discuss the implications of  these criteria.
Figure \ref{fig:graph_trace} shows the 90\% CL of $\Delta$ (top panels), $d_c$ (middle panels)
and $\gamma$ (bottom panels) as functions of the baryon density represented by blue (NJL) and
orange (MFTQCD) bands.
Concerning the trace anomaly, as discussed in \cite{Annala:2023cwx}, the pQCD constraints favors a positive $\Delta$.
Here, we identify some EOS for which $\Delta$ is always positive. 

\begin{figure}[h!]
	  \centering
	  \includegraphics[width=.95\linewidth]{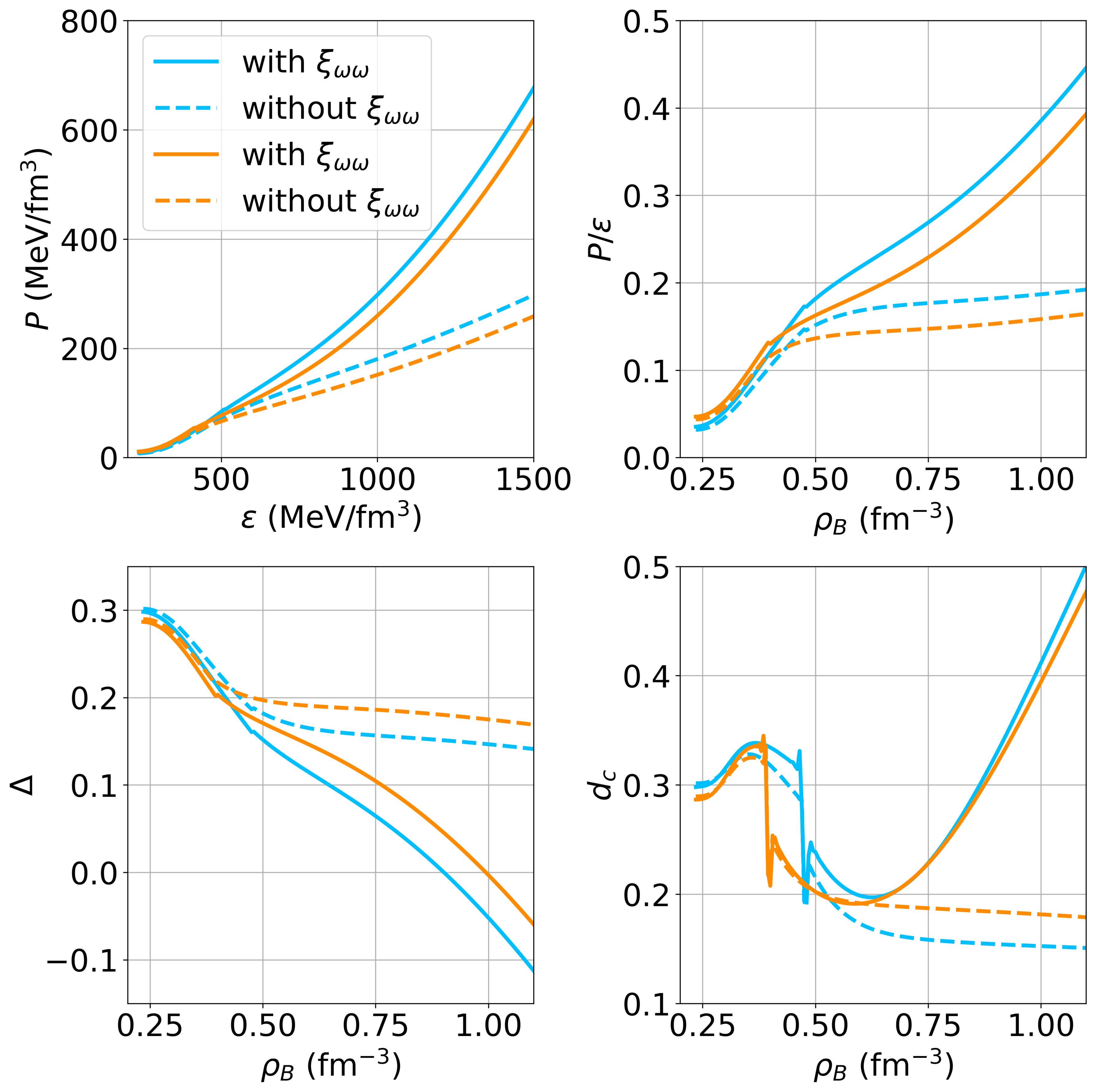}
	  \caption{Two EOSs from the NJL stiff set with pQCD constraints are shown as solid lines.  The parameter values are given in the table \ref{tab:parameters_fig}.  The dashed lines show the same equations but with $\xi_{\omega\omega} = 0$.}
	  \label{fig:referee_item4_2}
	\end{figure}

    It is interesting to understand the reason for the different behavior of $d_c$ in the two models. This can be attributed to the $\xi_{\omega\omega}$ term of the NJL model, which dominates at larger densities and contributes to a larger $P/\epsilon$ value. 
    In Fig. \ref{fig:referee_item4_2}, where two specific EOSs of the NJL stiff set are plotted with and without the $\xi_{\omega\omega}$ term (parameter values in table \ref{tab:parameters_fig}), this is clearly seen:
    the $\xi_{\omega\omega}$ term gives rise to much smaller (and negative) $\Delta$ values and $d_c$ only touches the 0.2, contrary to the behavior without this term, where the line $d_c=0.2$ is clearly crossed.

	\begin{table}[h!]
		\setlength{\tabcolsep}{7.pt}
		\renewcommand{\arraystretch}{1.2}
		\begin{tabular}{ccccccc}
			\hline \hline 
            color & $\xi_\omega$ & $\xi_\rho$ & $\xi_{\omega\omega}$ & $\xi_{\omega\sigma}$ & $\xi_{\omega\rho}$ & B \\
            \hline
			blue   & 0.145 & 0.070 & 16.49 & 0.789 & 5.300 & 7.664 \\
			orange & 0.019 & 0.274 & 14.27 & 1.825 & 17.73 & 9.761 \\
			\hline
		\end{tabular}
		\caption{Parameters for the two EOSs from the NJL stiff set with pQCD constraints plotted in Fig. \ref{fig:referee_item4_2}.}
		\label{tab:parameters_fig}
	\end{table}

\begin{figure}[!ht]
  \centering
    \includegraphics[width=.95\linewidth]{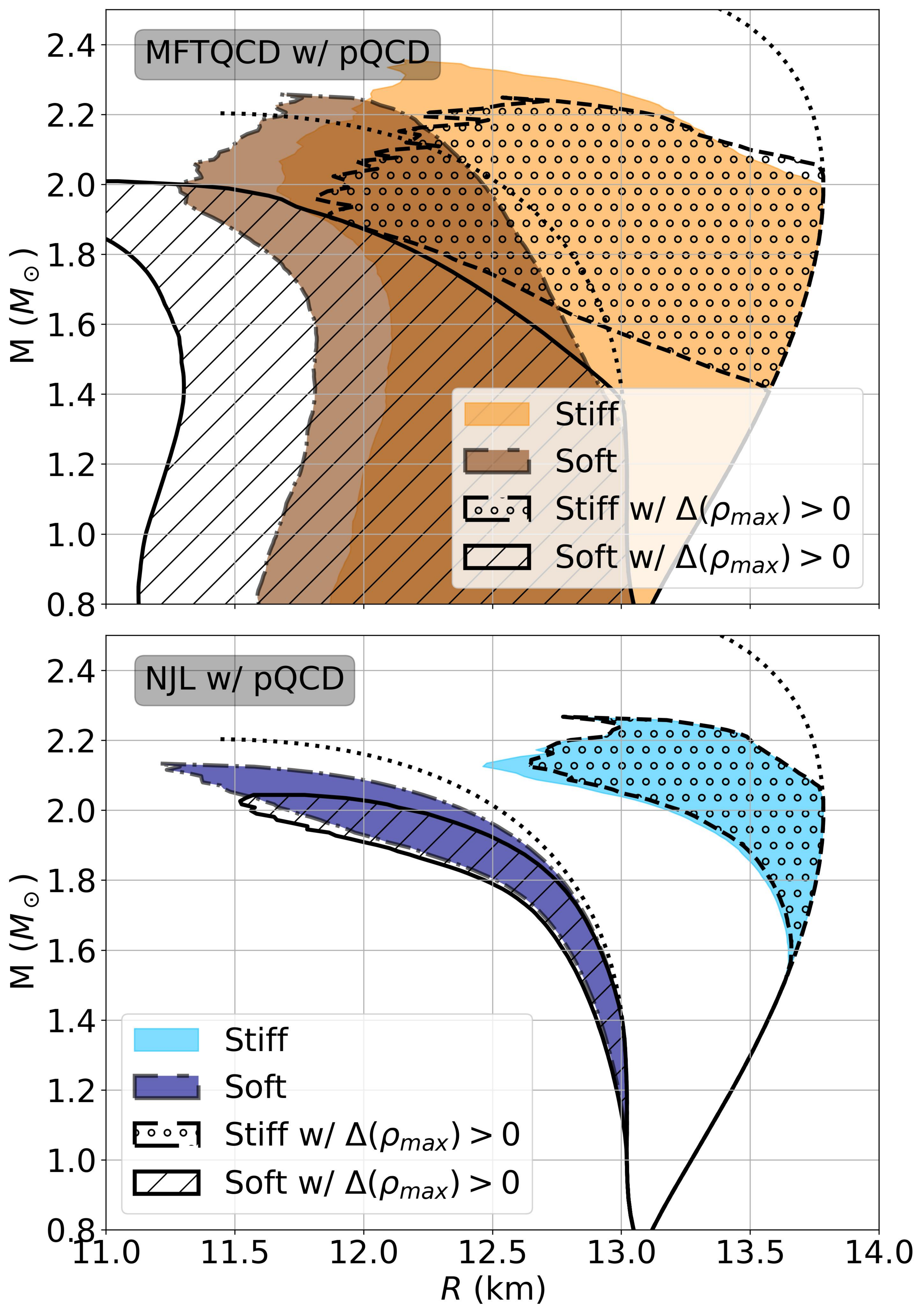}
  \caption{Mass-radius diagram of the sets with $\Delta(\rho_\text{max}) > 0$ filter (patterned bands).
  Sets without this filter are also shown in solid bands.
  Graphs on the left show the equations with the MFTQCD model, and on the right, with the NJL.}
  \label{fig:graph_TOV_Delta_p}
\end{figure}

\begin{figure*}
  \centering
  \begin{tabular}{cc}
    {\includegraphics[width=.45\linewidth]{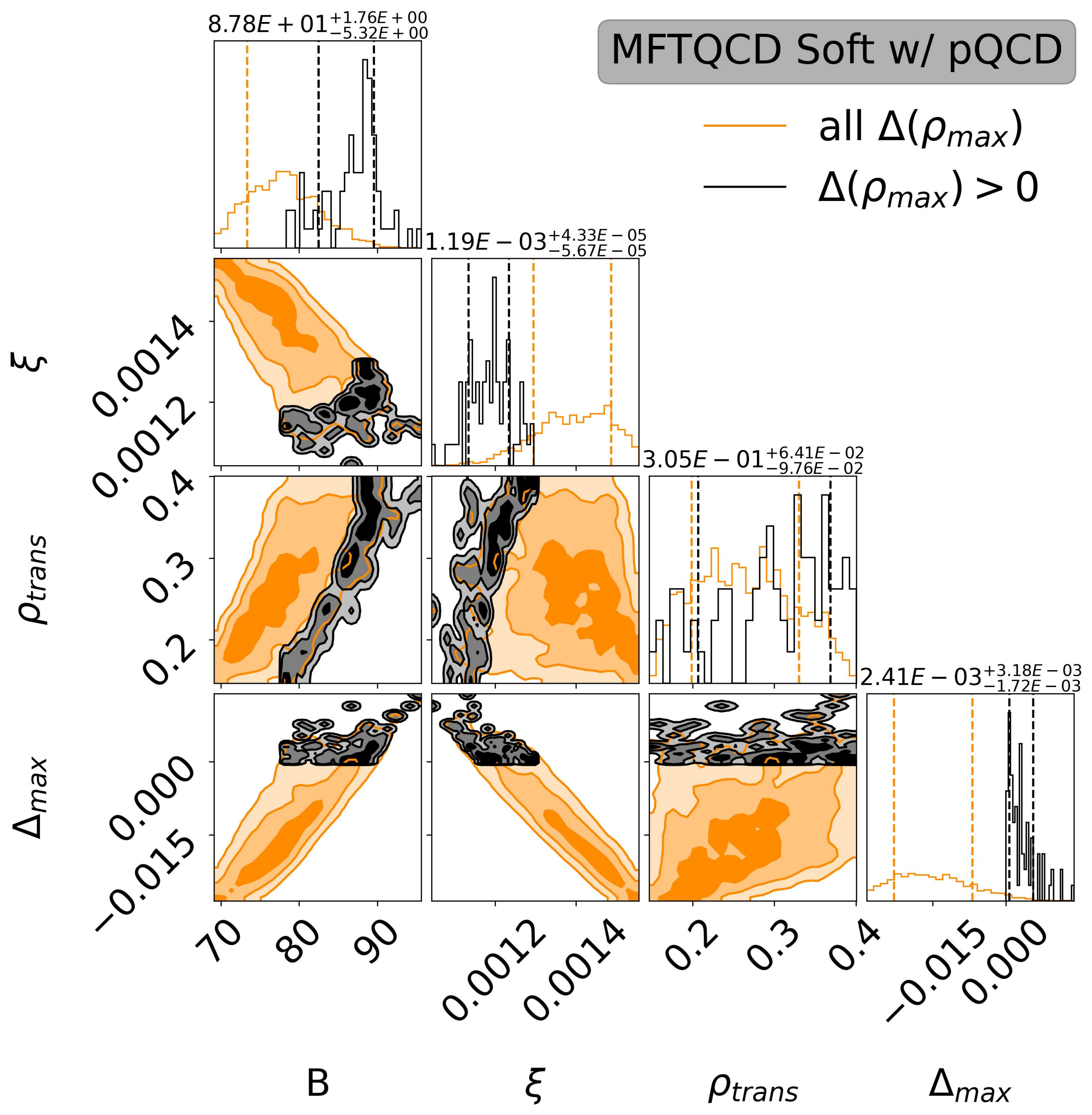}}
    &
    {\includegraphics[width=.45\linewidth]{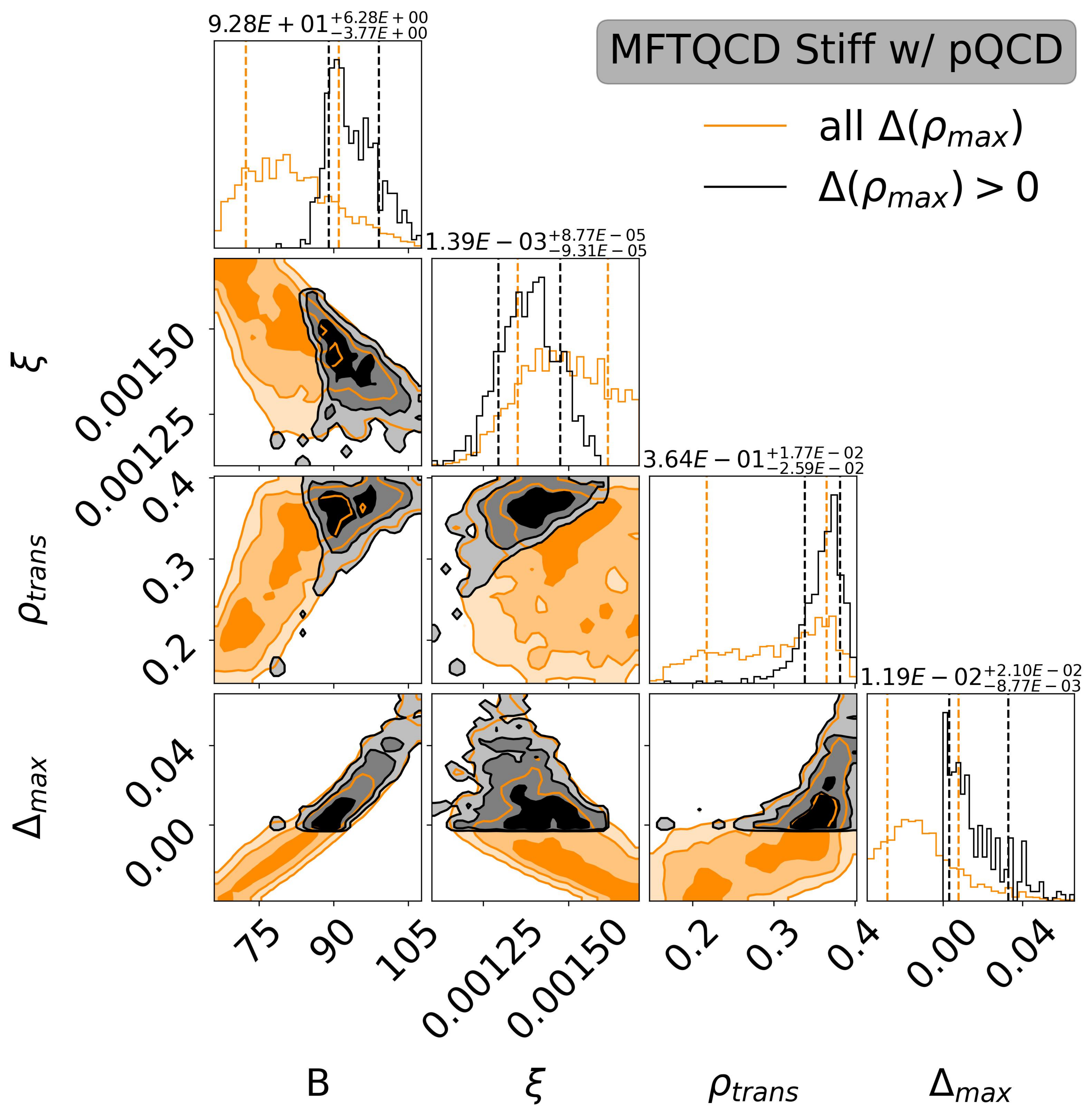}}
  \end{tabular}
  \caption{Corners of MFTQCD soft (left) and stiff (right).
    Orange (black) represents the sets without (with) the $\Delta(\rho_\text{max}) > 0$ filter.}
  \label{fig:corner_Delta_p}
\end{figure*}

 In Fig.
\ref{fig:graph_trace}, the 90\% CL of the $\Delta > 0$ at the maximum mass density ($\Delta_\text{max}$)
EOSs are represented by dark blue (NJL) and dark orange (MFTQCD) patterns.
NJL soft (stiff) EOS with $\Delta_\text{max} > 0$ represent 55\% (97\%) of the respective sets.
For MFTQCD soft (stiff) equations, they represent only 5\% (30\%).
The high percentage of EOS satisfying $\Delta_\text{max} > 0$
in the NJL stiff set is due to the fact that we have low $\rho_\text{max}$ values
(0.607 to 0.891 fm$^{-3}$ with 90\% CL - see Table \ref{tab:all}).
In Fig. \ref{fig:graph_trace}, we see that the band of EOS without the
$\Delta_\text{max} > 0$ filter already mostly satisfies this restriction.
It is also interesting to note that, when applying the $\Delta_\text{max} > 0$ filter,
we have a notable change in the phase transition values of the 90\% band in the MFTQCD stiff set.
This is because the MFTQCD stiff set is able to reach higher values of $B$.
From Eq. \ref{eq:mqcd},
we see that this parameter directly increases the $\Delta$ values.
However, $B$ also increases the phase transition values,
thus creating a correlation between $\Delta$ and $\rho_\text{trans}$.
These correlations can be seen in Fig. \ref{fig:corner_Delta_p},
 which shows the corner plots of the MFTQCD soft (left) and stiff (right) sets.
The effect of $\Delta_\text{max} > 0$ filter in the mass-radius diagram is shown in
Fig. \ref{fig:graph_TOV_Delta_p}.
Patterned (solid) bands show the equations with (without) the $\Delta_\text{max} > 0$ filter.
In article \cite{Fujimoto:2022ohj},
the authors analyzed the trace anomaly using agnostic equations.
When applying $\Delta > 0$, they observed a decrease in maximum mass and  its radius.
However, here this behavior can only be seen in the MFTQCD stiff and NJL soft sets.

Concerning $d_c$ and $\gamma$,
we verify that for all sets of models, the phase transition occurs before the graphs
reach the $d_c < 0.2$ and the $\gamma < 1.75$ region,
but rapidly attain a value below these.
After crossing the $d_c < 0.2$ value, the MFTQCD EOS show a monotonically decreasing behavior but the same is not true to NJL: the 8-quark term with the $\xi_{\omega\omega}$ coupling becomes stronger with density and gives rise to a non-monotonic behavior of $d_c$. At large densities $d_c$  rises above 0.2.
A similar behaviour can be seen in the $\gamma$ values.

\begin{table*}[]
\setlength{\tabcolsep}{6.pt}
\renewcommand{\arraystretch}{1.2}
\begin{tabular}{ccccccccccccc}
\hline \hline 
\multicolumn{13}{c}{\textbf{NJL}} \\ 
    &   \multicolumn{3}{c}{\textbf{soft w/ pQCD}}   &   \multicolumn{3}{c}{\textbf{soft w/o pQCD}} & \multicolumn{3}{c}{\textbf{stiff w/ pQCD}} & \multicolumn{3}{c}{\textbf{stiff w/o pQCD}}  \\
    &   \textbf{median} &   \textbf{min}    & \textbf{max}    &   \textbf{median} &   \textbf{min}    & \textbf{max} &   \textbf{median} &   \textbf{min}    & \textbf{max} &   \textbf{median} &   \textbf{min}    & \textbf{max} \\ \hline
$M_\text{max}$                  & 2.006 & 1.880 & 2.108& 1.983 & 1.872 & 2.074 & 2.082 & 1.980 & 2.175 & 2.134 & 2.001 & 2.335  \\
$\rho_\text{trans}$             & 0.358 & 0.322 & 0.389& 0.358 & 0.323 & 0.389& 0.374 & 0.323 & 0.399& 0.369 & 0.318 & 0.397 \\
$R_\text{max}$                  & 11.38 & 11.14 & 11.91& 11.45 & 11.17 & 11.95 & 13.00 & 12.03 & 13.62 & 12.49 & 11.71 & 13.54  \\
$P_0 (M_\text{max})$            & 399 & 243 & 526& 372 & 235 & 496& 184 & 106 & 350& 277 & 118 & 495 \\
$\epsilon_0 (M_\text{max})$     & 1237  & 1075  & 1304& 1216  & 1064  & 1300& 850 & 673 & 1092& 978 & 703 & 1174 \\
$\rho_0 (M_\text{max})$         & 1.000 & 0.912 & 1.031& 0.992 & 0.905 & 1.032 & 0.733 & 0.607 & 0.891 & 0.810 & 0.630 & 0.929  \\
$c^2_{s_0} (M_\text{max})$      & 0.626 & 0.340 & 0.832& 0.577 & 0.325 & 0.780& 0.323 & 0.190 & 0.651& 0.533 & 0.210 & 0.920 \\
$R (2 M_\odot)$                 & 11.81 & 11.38 & 12.05& 11.60 & 11.31 & 11.88 & 13.43 & 12.74 & 13.75 & 13.47 & 12.86 & 13.73  \\
$P_0 (2 M_\odot)$               & 215 & 184 & 307& 233 & 197 & 327 & 95  & 82  & 134 & 95  & 82  & 131  \\
$\epsilon_0 (2 M_\odot)$        & 839 & 736 & 1070& 890 & 782 & 1109& 525 & 396 & 677& 504 & 397 & 652 \\
$\rho_0 (2 M_\odot)$            & 0.844 & 0.770 & 0.968& 0.905 & 0.827 & 0.987 & 0.572 & 0.474 & 0.730 & 0.535 & 0.456 & 0.686  \\
$c^2_{s_0} (2 M_\odot)$         & 0.589 & 0.534 & 0.720& 0.649 & 0.571 & 0.749 & 0.294 & 0.212 & 0.447 & 0.353 & 0.238 & 0.453  \\
$R (1.8 M_\odot)$               & 12.69 & 12.51 & 12.77& 12.67 & 12.50 & 12.75& 13.76 & 13.56 & 13.76& 13.76 & 13.53 & 13.76 \\
$P_0 (1.8 M_\odot)$             & 114 & 108 & 130& 116 & 109 & 132& 63  & 63  & 75& 63  & 63  & 76 \\
$\epsilon_0 (1.8 M_\odot)$      & 582 & 544 & 670& 593 & 555 & 680& 363 & 358 & 451& 364 & 358 & 451 \\
$\rho_0 (1.8 M_\odot)$          & 0.547 & 0.517 & 0.618& 0.556 & 0.525 & 0.626 & 0.359 & 0.354 & 0.434 & 0.359 & 0.355 & 0.435  \\
$c^2_{s_0} (1.8 M_\odot)$       & 0.325 & 0.252 & 0.381& 0.317 & 0.247 & 0.359& 0.403 & 0.228 & 0.501& 0.372 & 0.240 & 0.499 \\
$R (1.4 M_\odot)$               & 12.98 & 12.92 & 13.00& 12.98 & 12.92 & 13.00& 13.57 & 13.57 & 13.57& 13.57 & 13.57 & 13.57 \\
$P_0 (1.4 M_\odot)$             & 53  & 52  & 55& 53  & 52  & 55& 37  & 37  & 37& 37  & 37  & 37 \\
$\epsilon_0 (1.4 M_\odot)$      & 409 & 402 & 422& 411 & 403 & 424& 306 & 306 & 306& 306 & 306 & 307 \\
$\rho_0 (1.4 M_\odot)$          & 0.405 & 0.398 & 0.417& 0.406 & 0.399 & 0.418 & 0.309 & 0.309 & 0.309 & 0.309 & 0.309 & 0.310  \\
$c^2_{s_0} (1.4 M_\odot)$       & 0.340 & 0.290 & 0.360& 0.338 & 0.283 & 0.358& 0.396 & 0.284 & 0.409& 0.393 & 0.270 & 0.409 \\
\hline \hline 
\multicolumn{13}{c}{\textbf{MFTQCD}} \\ 
    &   \multicolumn{3}{c}{\textbf{soft w/ pQCD}}   &   \multicolumn{3}{c}{\textbf{soft w/o pQCD}} & \multicolumn{3}{c}{\textbf{stiff w/ pQCD}} & \multicolumn{3}{c}{\textbf{stiff w/o pQCD}}  \\
    &   \textbf{median} &   \textbf{min}    & \textbf{max}    &   \textbf{median} &   \textbf{min}    & \textbf{max} &   \textbf{median} &   \textbf{min}    & \textbf{max} &   \textbf{median} &   \textbf{min}    & \textbf{max} \\ \hline
$M_\text{max}$                  & 2.101 & 1.968 & 2.218 & 2.105 & 1.957 & 2.212 & 2.143 & 1.955 & 2.318 & 2.161 & 1.965 & 2.378 \\
$\rho_\text{trans}$             & 0.259 & 0.173 & 0.365 & 0.256 & 0.174 & 0.369 & 0.316 & 0.185 & 0.388 & 0.311 & 0.183 & 0.388 \\
$R_\text{max}$                  & 11.27 & 10.78 & 11.61  & 11.27 & 10.77 & 11.60 & 11.80 & 11.05 & 13.32 & 11.86 & 11.07 & 12.74 \\
$P_0 (M_\text{max})$            & 432 & 408 & 459  & 432 & 410 & 459 & 396 & 88  & 445 & 393 & 89  & 444 \\
$\epsilon_0 (M_\text{max})$     & 1239  & 1135  & 1370 & 1237  & 1140  & 1378 & 1135  & 791 & 1312 & 1126  & 802 & 1303 \\
$\rho_0 (M_\text{max})$         & 0.993 & 0.915 & 1.100 & 0.991 & 0.919 & 1.104 & 0.913 & 0.706 & 1.050 & 0.906 & 0.712 & 1.043 \\
$c^2_{s_0} (M_\text{max})$      & 0.488 & 0.465 & 0.503  & 0.488 & 0.463 & 0.503 & 0.495 & 0.355 & 0.521 & 0.498 & 0.366 & 0.528 \\
$R (2 M_\odot)$                 & 12.01 & 11.43 & 12.37 & 12.00 & 11.46 & 12.35 & 12.68 & 11.72 & 13.49 & 12.77 & 11.78 & 13.57 \\
$P_0 (2 M_\odot)$               & 199 & 148 & 339 & 195 & 150 & 321 & 145 & 86  & 275 & 138 & 86  & 272 \\
$\epsilon_0 (2 M_\odot)$        & 742 & 598 & 1086 & 732 & 602 & 1047 & 621 & 483 & 947 & 605 & 447 & 940 \\
$\rho_0 (2 M_\odot)$            & 0.714 & 0.601 & 0.908 & 0.711 & 0.609 & 0.897 & 0.613 & 0.502 & 0.844 & 0.586 & 0.464 & 0.830 \\
$c^2_{s_0} (2 M_\odot)$         & 0.471 & 0.465 & 0.479 & 0.471 & 0.465 & 0.478 & 0.473 & 0.463 & 0.478 & 0.473 & 0.464 & 0.481 \\
$R (1.8 M_\odot)$               & 12.28 & 11.64 & 12.64 & 12.28 & 11.61 & 12.66 & 13.17 & 12.00 & 13.76 &  13.19 & 12.03 & 13.76 \\
$P_0 (1.8 M_\odot)$             & 126 & 103 & 175 & 125 & 104 & 179 & 88  & 63  & 148 & 86  & 63  & 145 \\
$\epsilon_0 (1.8 M_\odot)$      & 586 & 499 & 742 & 583 & 503 & 756 & 483 & 362 & 668 & 475 & 363 & 661 \\
$\rho_0 (1.8 M_\odot)$          & 0.555 & 0.484 & 0.683 & 0.553 & 0.487 & 0.693 & 0.465 & 0.358 & 0.621 & 0.457 & 0.358 & 0.615 \\
$c^2_{s_0} (1.8 M_\odot)$       & 0.443 & 0.432 & 0.451 & 0.443 & 0.430 & 0.451 & 0.447 & 0.363 & 0.497 & 0.449 & 0.385 & 0.497 \\
$R (1.4 M_\odot)$               & 12.42 & 11.81 & 12.93 & 12.41 & 11.78 & 12.94 & 13.57 & 12.09 & 13.57 & 13.57 & 12.20 & 13.57 \\
$P_0 (1.4 M_\odot)$             & 64  & 55  & 79 & 63  & 55  & 80 & 37  & 37  & 69 & 38  & 37  & 68 \\
$\epsilon_0 (1.4 M_\odot)$      & 441 & 399 & 508 & 441 & 400 & 511 & 307 & 306 & 467 & 329 & 306 & 468 \\
$\rho_0 (1.4 M_\odot)$          & 0.437 & 0.400 & 0.498 & 0.436 & 0.401 & 0.502 & 0.310 & 0.309 & 0.461 & 0.332 & 0.309 & 0.460 \\
$c^2_{s_0} (1.4 M_\odot)$       & 0.426 & 0.407 & 0.437 & 0.426 & 0.407 & 0.437 & 0.399 & 0.309 & 0.447 & 0.401 & 0.295 & 0.451 \\
\hline
\end{tabular}
\caption{The median and the 90\% CL of NS properties for all sets:
maximum mass $M_\text{max}$ ($M_\odot$);
transition baryon density $\rho_\text{trans}$ (fm$^{-3}$);
radium $R_\text{i}$ (km),
central pressure $P_{0, i}$ (MeV/fm$^{-3}$),
central energy density $\epsilon_{0, i}$ (MeV/fm$^{-3}$), 
central baryonic density $\rho_{0, i}$ (fm$^{-3}$)
at $M_i$, with $i = [ \text{max}, 2 M_\odot, 1.8 M_\odot, 1.4 M_\odot ]$.
}
\label{tab:all}
\end{table*}

\section{Conclusions}\label{chap:conclusion}
In the present study we have as main objective to understand the possibility that NS have a deconfined quark phase, considering the quark phase described by  phenomenological microscopic models, and avoiding agnostic descriptions.
For the hadron phase, we use two extreme  hadron EOS from \cite{constrain_apply}:
the BMPF 220 (soft) and the BMPF 260 (stiff) that satisfy nuclear matter properties and observational constraints.
For the quark phase, we consider the chiral symmetric SU(3)  Nambu-Jona-Lasinio (NJL) model
with 4-quark and 8-quark interaction terms, and  the Mean Field Theory of QCD (MFTQCD) which results from the QCD Lagrangian, considering a  gluon field decomposition in soft and 
hard momentum components \cite{Fogaca:2010mf, Franzon:2012in, Albino:2021zml}.

Based on these models we have  generated eight different sets choosing two hadron EOS and the two quark model, and  considering/not considering  the very high density pQCD constraints.
 Bayesian inference is applied to determine the quark models  coupling constants and a constant bag term $B$. We have imposed as constraints the observational data from
PSR J0030+0451 \cite{Riley:2019yda, Miller:2019cac} and
PSR J0740+6620 \cite{Riley:2021pdl, Miller:2021qha}.
Besides, we have also imposed that  the phase transition should  occur around 
$\rho_\text{trans} \approx 0.15 - 0.4 \text{ fm}^{-3}$.
In four  sets, we have also enforced the pQCD constraint at $\rho_B = 7 n_0$ (where $n_0=0.16$ fm$^{-3}$)
\cite{Komoltsev2021jzg}.
We have confirmed that the current observational data are compatible with the existence of a quark core inside NS. The pQCD constraints were important to constrain the NJL model to give causal EOS, since the  NJL 8-quark term which is responsible for allowing two solar mass stars with a large quark core may give rise to a super-luminous speed of sound. 
For both quark models we have obtained for the maximum star mass a value of the order of 2.1--2.3$M_\odot$ and similar radii. However, with MFTQCD quark model it was possible to obtain medium and low mass NS with a smaller radius, in particular, below 12 km (even compatible with the compact object HESS J1731-347 \cite{hess}). With MTFQCD we have obtained a transition to quark matter above $\sim 0.15$ fm$^{-3}$ but within NJL the transition density was above $\sim 0.25$ fm$^{-3}$. This explains why within NJL the medium and low mass stars have a larger radius. {Note that in \cite{Pfaff:2021kse}, only four quark interactions were introduced in the NJL Lagrangian density, and, as a consequence, quark cores were only found in the most massive stars. In \cite{Alvarez-Castillo:2016oln},  multiquark interactions were considered in the NJL model, however, the main objective of the study was to determine under which conditions a NS "third family" with a disconnected hybrid star branch  exists. In our analysis, the multiquark interactions considered do not predict a third family, since we did not include a eight quark scalar-pseudoscalar interaction because we did not want to affect the vacuum properties.} 

We have discussed the behavior of the normalized matter trace anomaly and the derived quantity $d_c$ proposed to measure conformality in \cite{Annala:2023cwx} and could conclude that NS may have a quark core but within the present description, quark matter is still strongly interacting matter and the  conformal limit is not attained in the NS center. {Besides the multiquark vector interactions may give rise to values of $d_c>0.2$ and $\gamma>1.7$ in a deconfined phase.} 

In a future work, we will apply the Bayesian inference also in the hadronic phase.
In that way, we will obtain all the range of the equations of state.
We also plan to compare to purely hadronic sets and analyze
if hybrid stars are more probable than hadronic stars based on the current observational data.

\section*{Acknowledgments}
M.A. expresses sincere gratitude to the FCT for their generous support through Ph.D. grant number 2022.11685.BD. This research received partial funding from national sources through FCT (Fundação para a Ciência e a Tecnologia, I.P, Portugal) for projects UIDB/04564/2020 and UIDP/04564/2020, identified by DOI 10.54499/UIDB/04564/2020 and 10.54499/UIDP/04564/2020, respectively, as well as for project 2022.06460.PTDC with DOI 10.54499/2022.06460.PTDC. The authors acknowledge the Laboratory for Advanced Computing at the University of Coimbra for providing {HPC} resources that have contributed to the research results reported within this paper, URL: \hyperlink{https://www.uc.pt/lca}{https://www.uc.pt/lca}.

\onecolumngrid

\appendix

\section{Parameter and phase transition posteriors}
\label{section:appendixA}
\begin{figure}[!ht]
  \centering
  \begin{tabular}{cc}
    {\includegraphics[width=.45\linewidth]{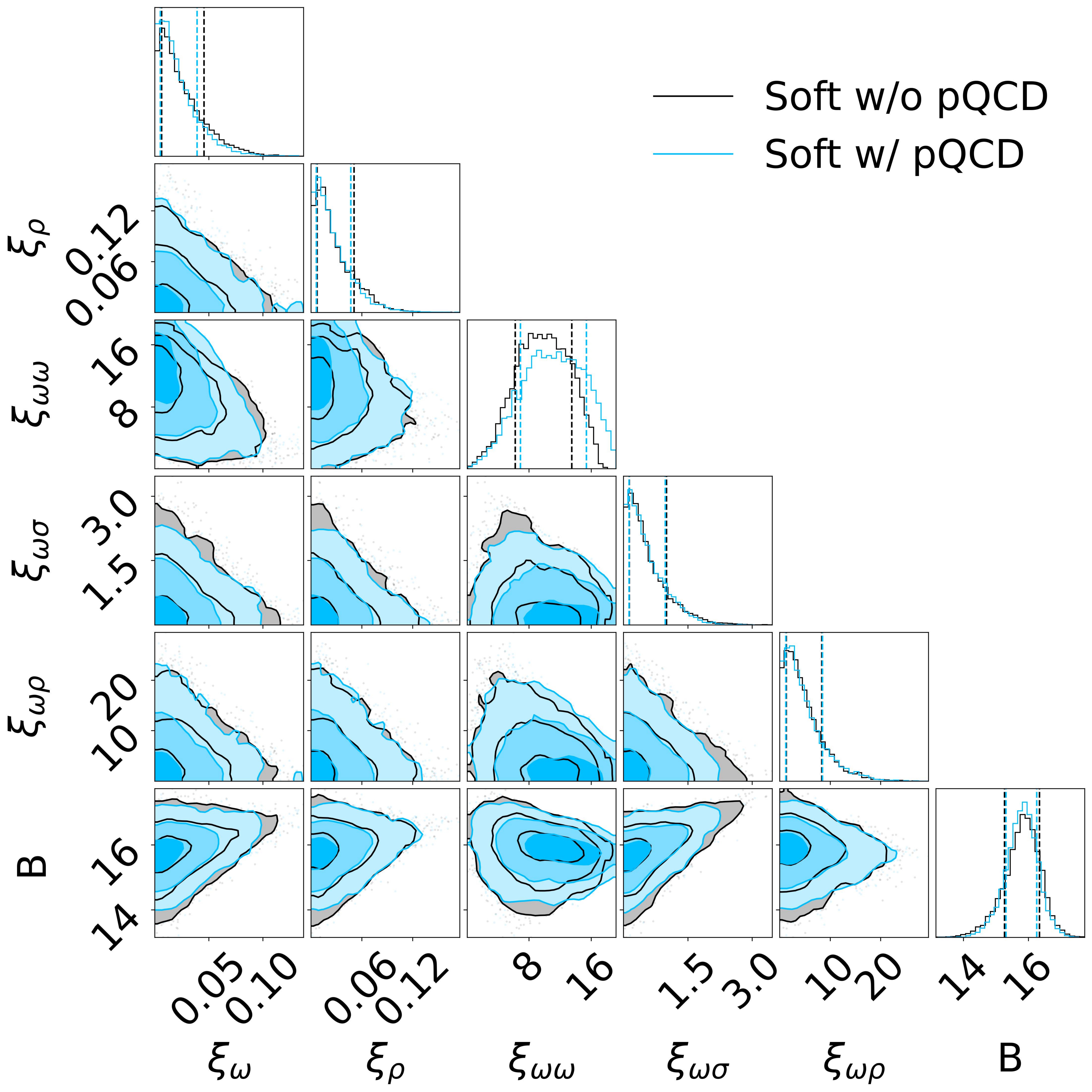}}
    &
    {\includegraphics[width=.45\linewidth]{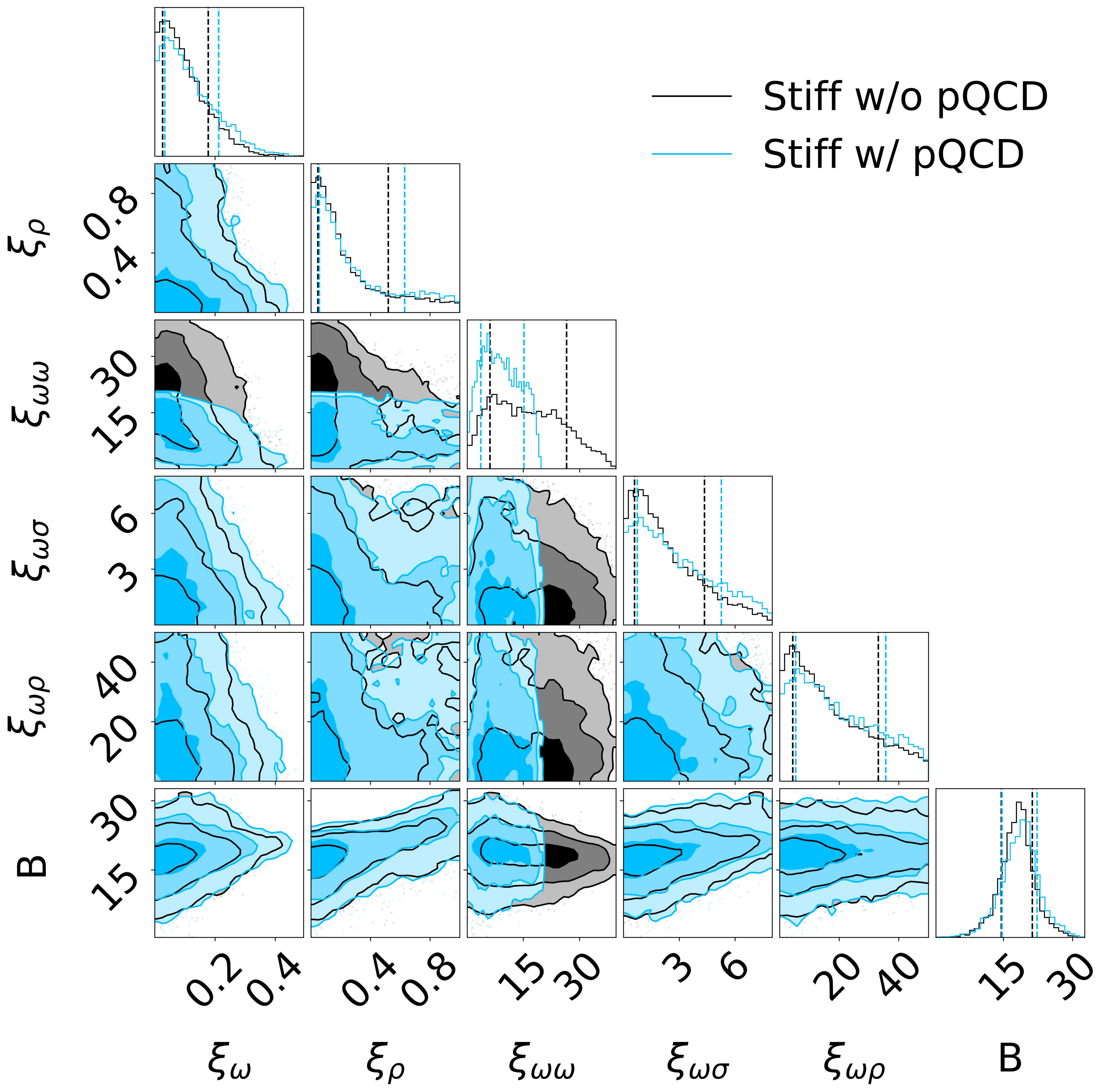}}
  \end{tabular}
  \caption{Corner plots of the parameters values of NJL soft (left) and NJL stiff (right).
  Models with pQCD constraints are plotted in blue
  and without pQCD in black.}
  \label{fig:corner_parameters_njl}
\end{figure}

\begin{figure}[!ht]
  \centering
  \begin{tabular}{cccc}
    {\includegraphics[width=.45\linewidth]{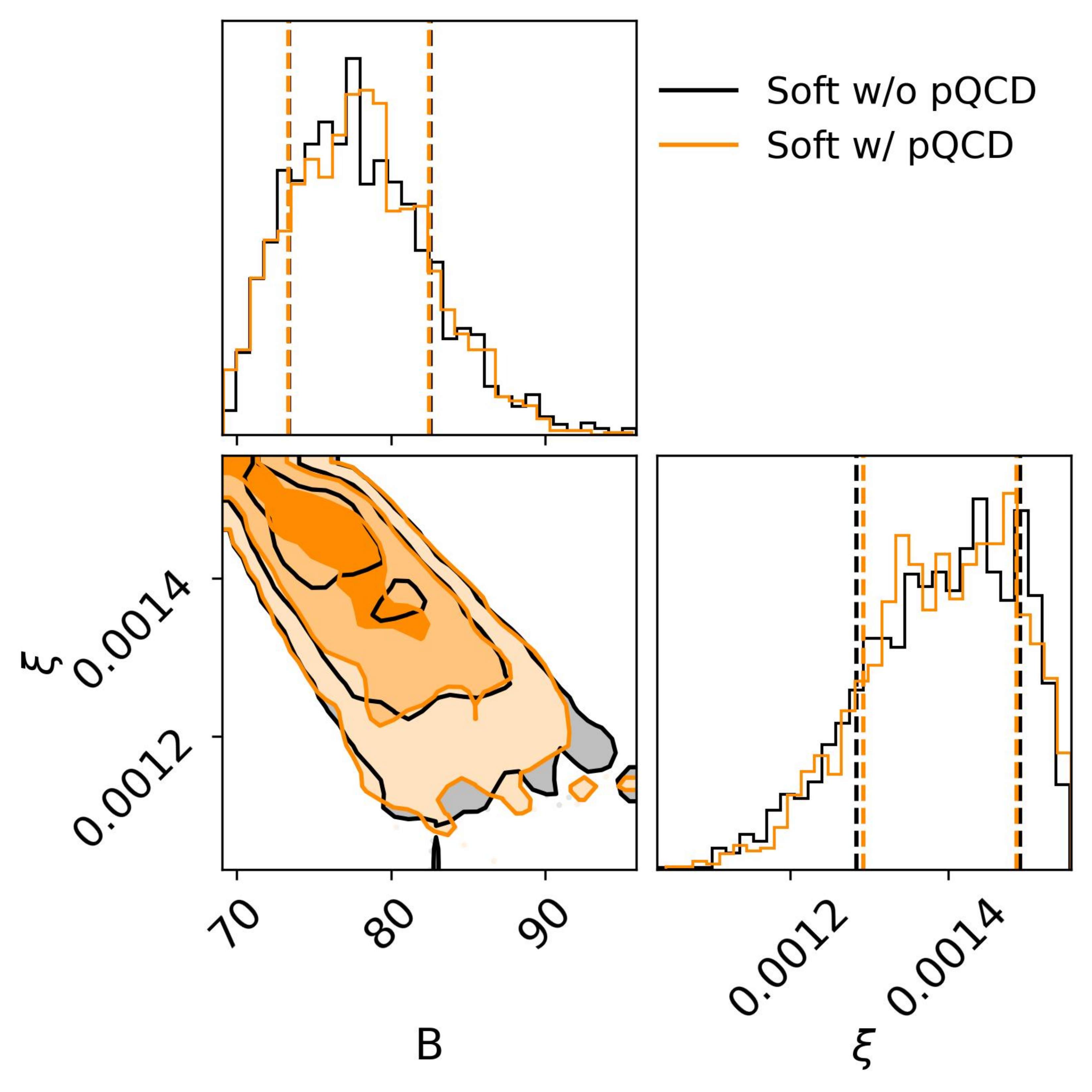}}
    &
    {\includegraphics[width=.45\linewidth]{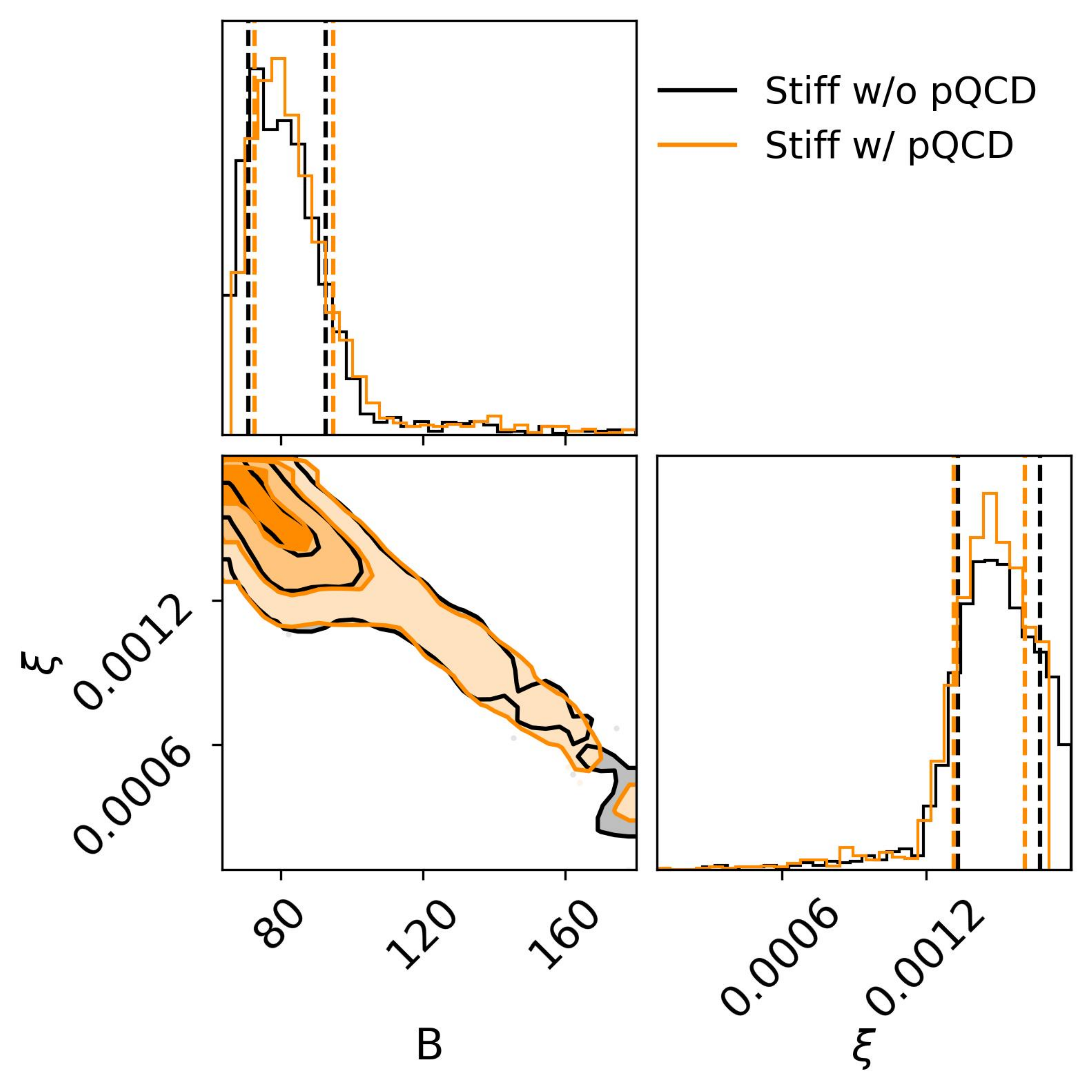}}
  \end{tabular}
  \caption{Corner plots of the parameters values of MFTQCD soft (left) and MFTQCD stiff (right).
  Models with pQCD constraints are plotted in orange
  and without pQCD in black.}
  \label{fig:corner_parameters_mqcd}
\end{figure}

\begin{figure}[!ht]
  \centering
  \begin{tabular}{cc}
    {\includegraphics[width=.45\linewidth]{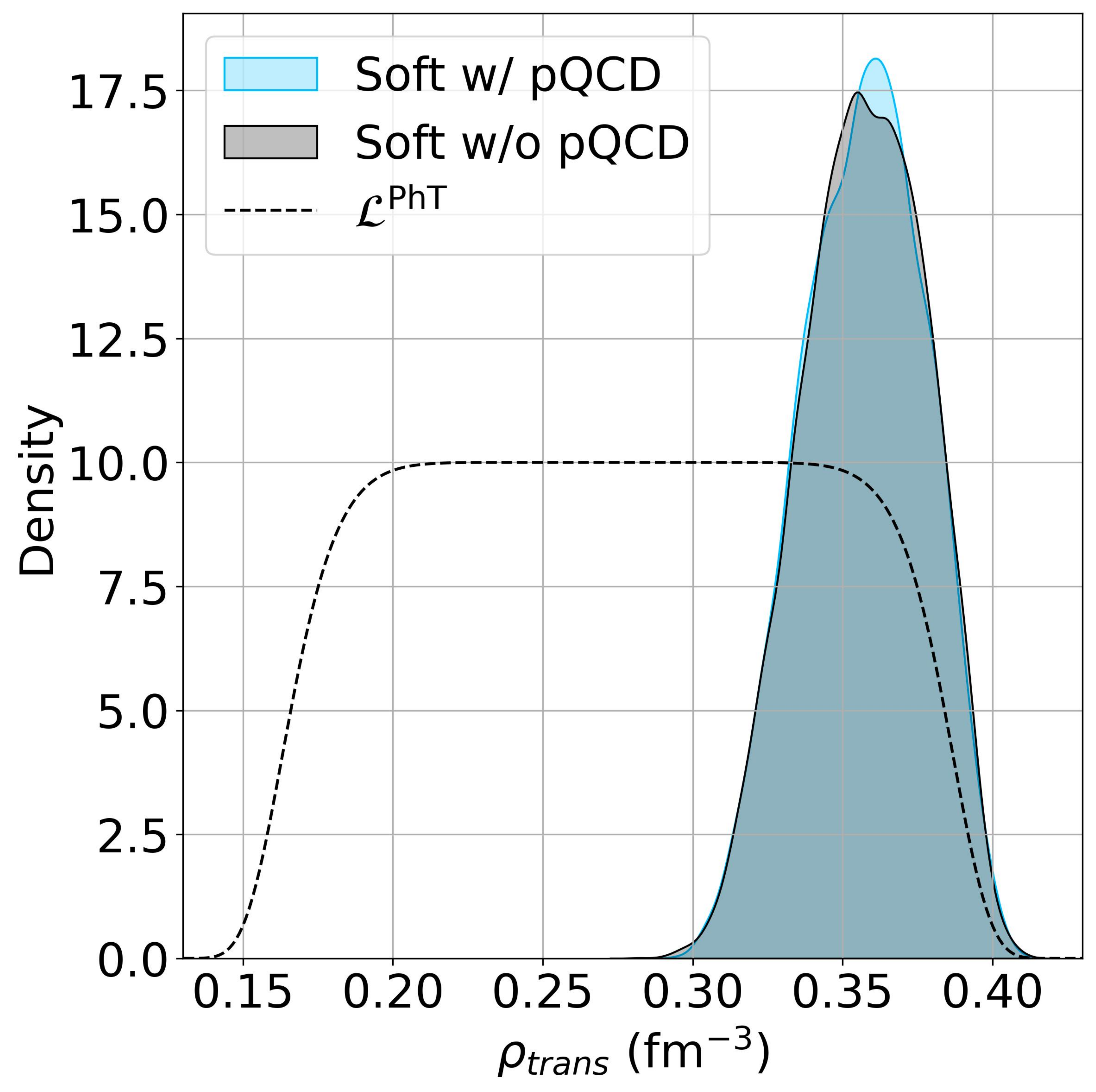}}
    &
    {\includegraphics[width=.45\linewidth]{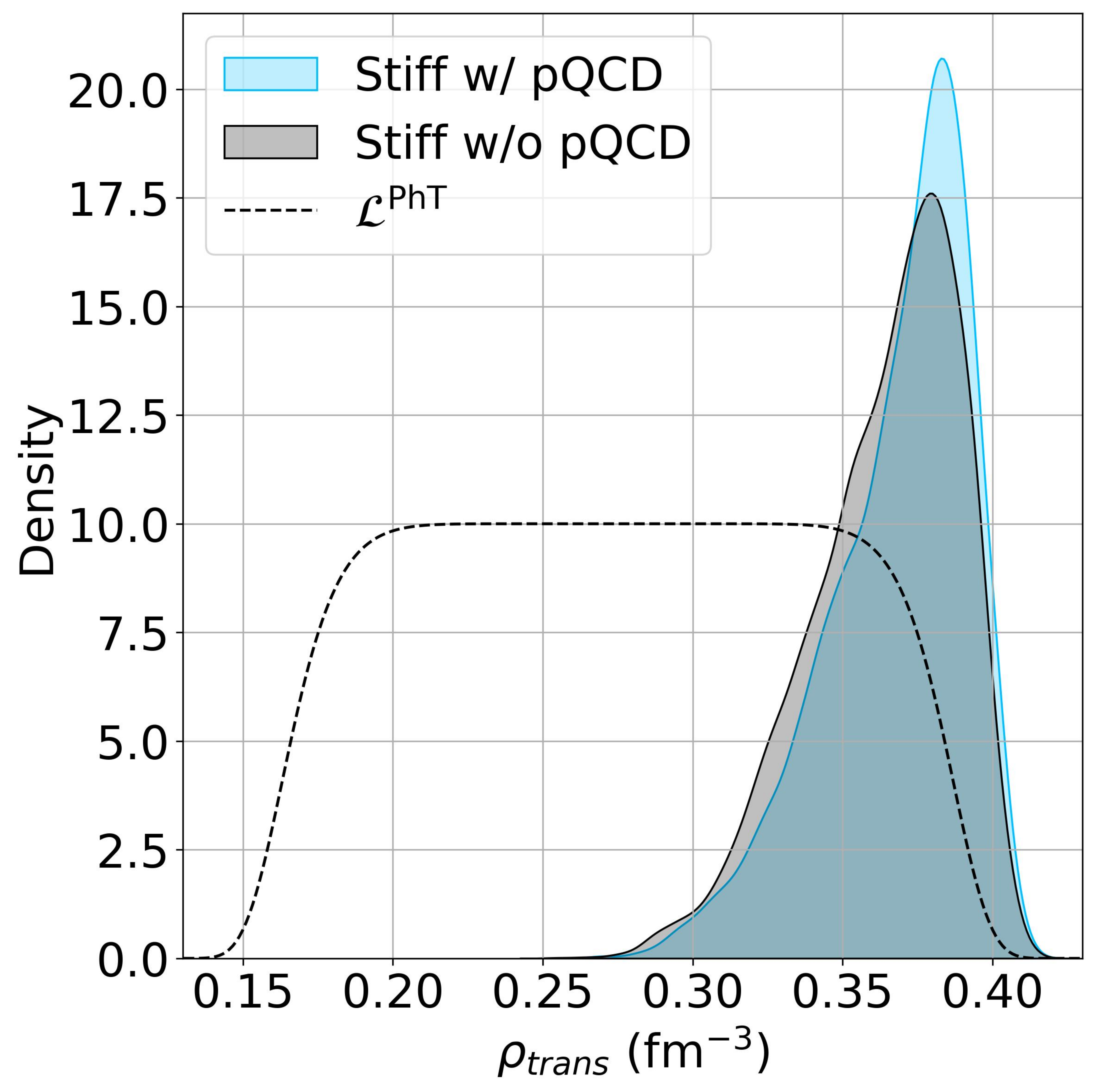}}
    \\
    {\includegraphics[width=.45\linewidth]{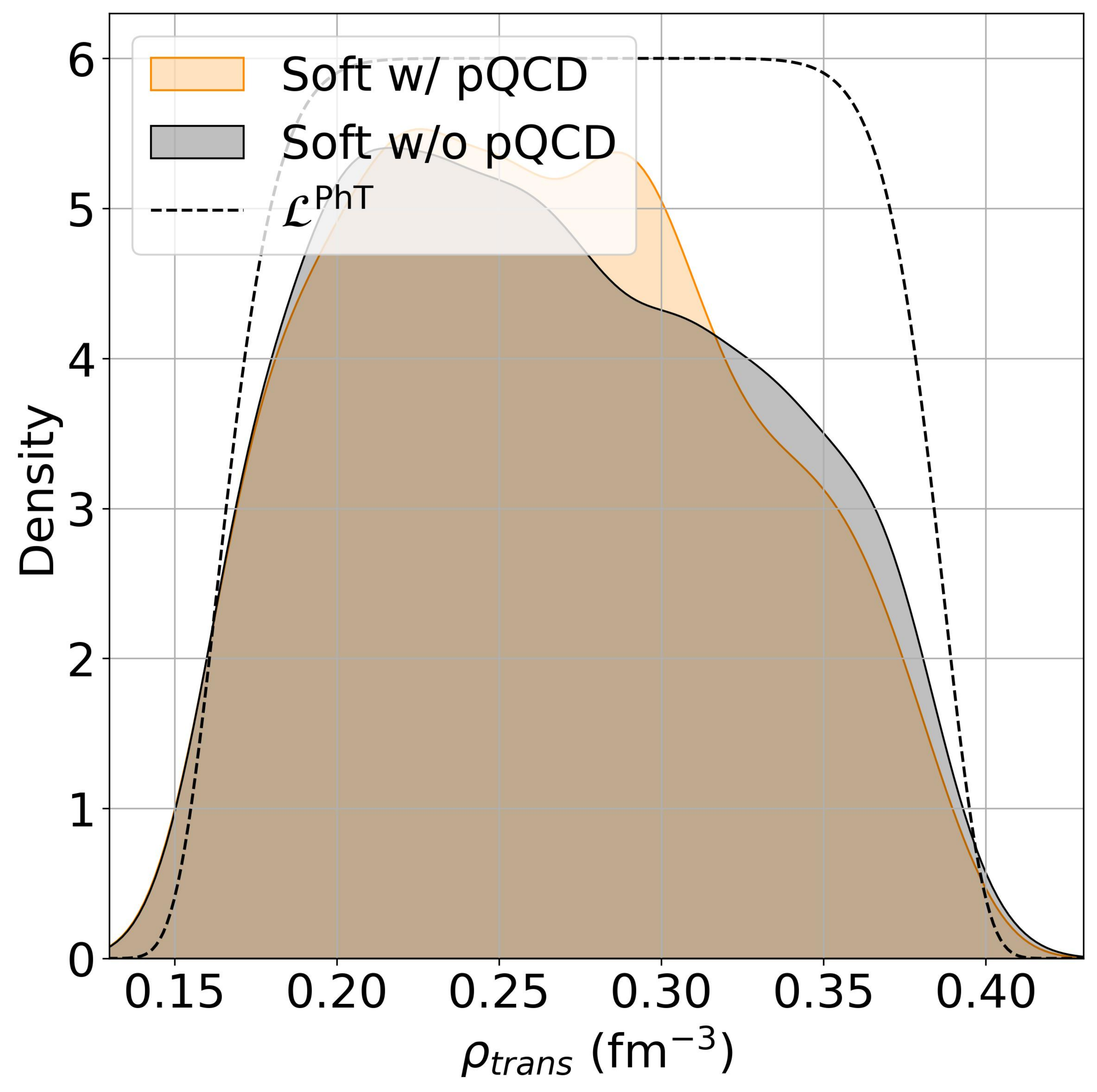}}
    &
    {\includegraphics[width=.45\linewidth]{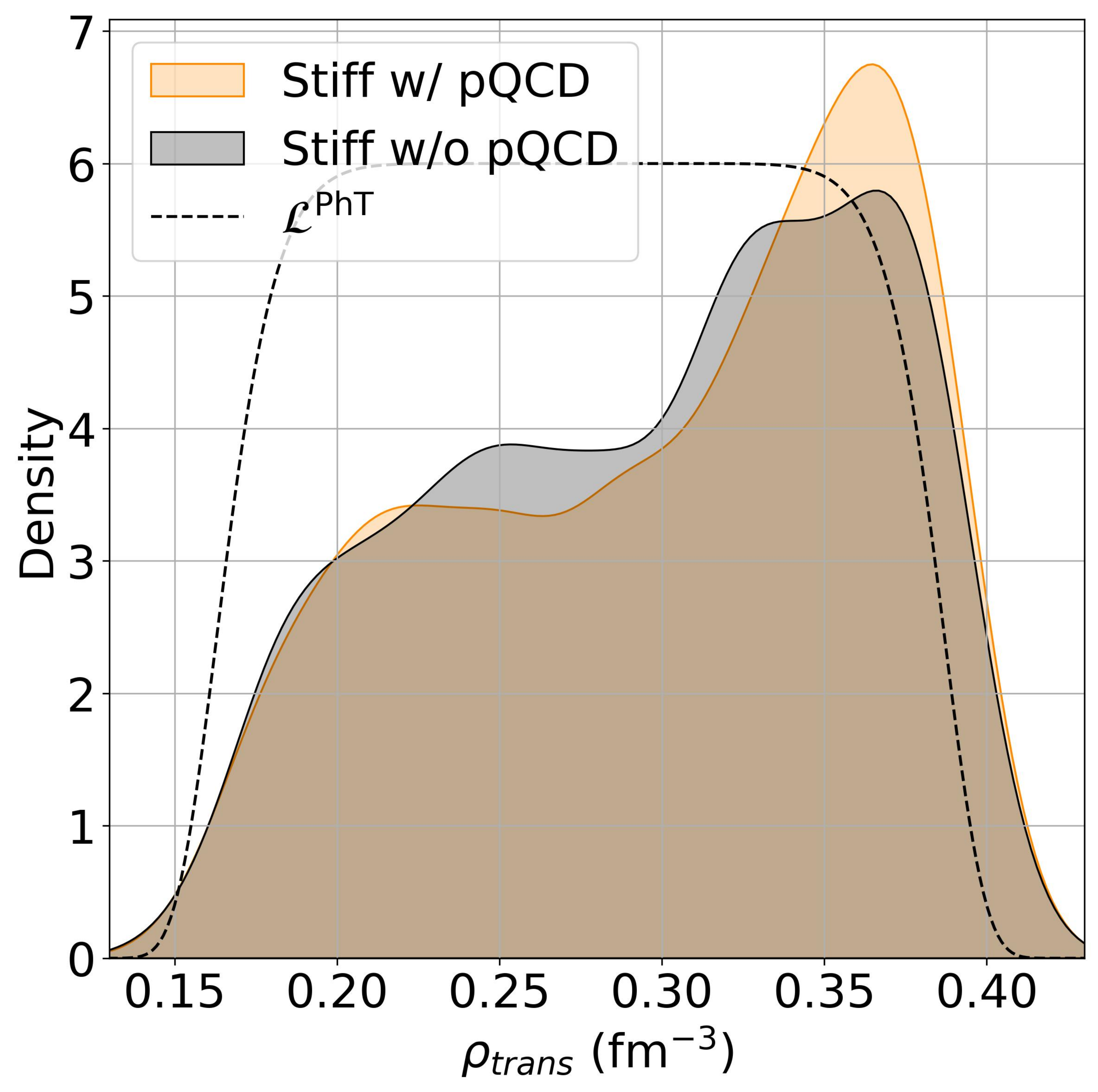}}
  \end{tabular}
  \caption{The phase transition likelihood ($\mathcal{L}^{\rm PhT}$) and the marginalized posteriors.
  The upper figures show the graphs for the NJL model and the lower ones for the MFTQCD model.
  On the left, we have the soft models and on the right, the stiff models.
  Models with pQCD constraints are plotted in blue (NJL) or orange (MFTQCD)
  and without pQCD in black.
  Dotted lines shows the imposed constraint in phase transition ($\mathcal{L}^\text{PhT}$).}
  \label{fig:likelihood}
\end{figure}

In this section, we present some additional results that support our discussion.
Figures \ref{fig:corner_parameters_njl} and \ref{fig:corner_parameters_mqcd}
show the corner plots of the NJL and MFTQCD models parameters, respectively.
Each corner shows a set with the pQCD constraint (blue or orange)
and the same set without the constraint (black).
Figure \ref{fig:likelihood} shows the  probability distribution of the baryonic density of phase transition.
Plots are arranged as shown in figure \ref{fig:corner_parameters_njl} and \ref{fig:corner_parameters_mqcd}.

\clearpage
\twocolumngrid

\bibliography{referencias}

\end{document}